\documentclass[%
 amsmath,amssymb,
twocolumn,
]{revtex4-1}

\usepackage{lineno}
\usepackage{graphicx}
\usepackage{dcolumn}
\usepackage{bm}
\usepackage{color}
\usepackage{wrapfig}
\usepackage{xcolor,colortbl}

\def\lapprox{\mathrel{\hbox{\rlap{\hbox{\lower4pt\hbox{$\sim$}}}\hbox{$<$}}}}
\def\gapprox{\mathrel{\hbox{\rlap{\hbox{\lower4pt\hbox{$\sim$}}}\hbox{$>$}}}}

\newcommand{\be}{\begin{equation}}
\newcommand{\ee}{\end{equation}}

\newcommand {\nind} {\noindent}
\newcommand {\mb} {\mathbf}

\newcommand {\bea} {\begin{eqnarray}}
\newcommand {\eea} {\end{eqnarray}}

\newcommand {\eps} {\epsilon}

\newcommand{\jl}{}

\begin{document}

\title{Testing the Accuracy of Data Driven MHD Simulations of Active Region Evolution}

\author{James E. Leake}
\email{james.e.leake@nasa.gov. JEL previously at [1], now at [2].}
\affiliation{ U.S. Naval Research Lab 4555 Overlook Ave., SW Washington, DC 20375}
\affiliation{NASA Goddard Space Flight Center, 8800 Greenbelt Rd, Greenbelt, MD 20771}
\author{Mark G. Linton}
\affiliation{ U.S. Naval Research Lab 4555 Overlook Ave., SW Washington, DC 20375}
\author{Peter W. Schuck}%
\affiliation{NASA Goddard Space Flight Center, 8800 Greenbelt Rd, Greenbelt, MD 20771}


\date{\today}

\begin{abstract}

{\jl Models for the evolution of the solar coronal magnetic field are vital for understanding solar activity, yet the best measurements of magnetic field lie at the photosphere, necessitating the development of coronal models which are \textit{``data-driven''} at the photosphere.  We present an investigation to determine the feasibility and accuracy of such methods. Our validation framework uses a simulation of active region (AR) formation, modeling the emergence of magnetic flux from the convection zone to the corona, as a ground-truth dataset, to supply both the photospheric information, and to perform the validation of the data-driven method. We focus our investigation on how the data-driven model accuracy depends on the temporal frequency of the driving data. The Helioseismic and Magnetic Imager on NASA's Solar {\jl Dynamics} Observatory produces full-disk vector magnetic field measurements at a 12 minute cadence. Using our framework we show that ARs that emerge over 25 hours can be modeled by the data-driving method with only $\sim$1\% error in the free magnetic energy, assuming the photospheric information is specified every 12 minutes. However, for rapidly evolving features, under-sampling of the dynamics at this cadence leads to a strobe effect, generating large electric currents and incorrect coronal morphology and energies. We derive a sampling condition for the driving cadence based on the evolution of these small-scale features, and show that higher-cadence driving can lead to acceptable errors. Future work will investigate the source of errors associated with deriving plasma variables from the photospheric magnetograms as well as other sources of errors such as reduced resolution and instrument bias and noise. }
\end{abstract}

\maketitle

\section{Introduction and Main Result}

In order to understand the complex interplay of magnetic, hydrodynamic, and thermal forces associated with the onset of various manifestations of solar activity, it is vital to know the magnetic field in the solar corona, where the storage and release of energy in non-potential ($\mathbf{J} = \nabla\times\mb{B}/\mu_{0} \ne 0$) magnetic fields is the cause of solar flares, coronal mass ejections (CMEs), and prominence eruptions. However, as yet no method exists to accurately achieve this goal. Fortunately, multiple missions such as the Solar Optical Telescope on Hinode \citep[SOT,][]{SOT_paper}, the Helioseismic and Magnetic Imager on the Solar Dynamics Observatory \citep[HMI/SDO,][]{HMI_paper}, Big Bear Solar Observatory's New Solar Telescope \citep{NST_paper}, and the National Solar Observatory's upcoming Daniel K. Inouye Solar Telescope \citep[DKIST,][]{DKIST_paper} are/will be able to observe the magnetic field at the photosphere. 
Currently, with HMI there are observations of the Sun's surface vector magnetic field at one optical depth every 12 minutes. These data can be used in a variety of ways to estimate the coronal magnetic field, using a number of assumptions. 

One approach is to use measurements of the magnetic field at the photospheric surface, assuming that the constant optical depth surface can be transformed to a constant depth surface, and extrapolate the magnetic field into the domain above this surface, using either the potential, $\mathbf{J} = 0$, or more general force-free,  $\mathbf{J}\times\mathbf{B}=0$ \citep[e.g.,][]{Wheatland_2000,Wiegelmann_2004}, approximations. These extrapolation methods are able to produce estimates of the 3D magnetic field at discrete times during an active region's evolution based on the magnetogram data, and are used to estimate useful indices, such as the free magnetic energy  ($E_{free} \equiv E-E_{pot}$ where $E$ is the total magnetic energy, and $E_{pot}$ is the energy of the potential magnetic field). 
However, the approximations used in these extrapolations (namely $\beta=2\mu_{0}P/B^{2}=0$ where $P$ and $B$ are the gas pressure and magnetic field strength, respectively) are not generally true in the lower solar atmosphere above where the photospheric data are observed.
Furthermore, \citet{Peter_2015} recently revisited the basic assumptions associated with the force-free field extrapolations, guided by analysis of the MHD equations, global energy considerations, and 3D MHD simulations. Their analysis suggests that the relative difference between the total energy in a volume of the atmosphere, and that obtained from a force-free extrapolation, which assumes $\beta=0$, can be of the order of the plasma $\beta$ averaged over that volume. 



In  recent  years, the dynamic coronal magnetic field has been modeled by considering the field evolution as a succession of equilibria, with boundary conditions informed by measurement of the photospheric magnetic field. These \textit{``data-driven''} models have various assumptions. In the magneto-frictional (MF) approach \citep[e.g.,][]{Cheung2012, Weinzierl2016}, the lower boundary of a zero-$\beta$ (no plasma forces) coronal model is periodically updated with photospheric magnetic field observations, and the velocity everywhere is set proportional to the Lorentz force. Then an equilibrium is found before the next input of photospheric  data. 

Relaxing the zero-$\beta$ approximation, recent data-driven MHD approaches assume a low $\beta$ atmosphere and update the lower boundary with the observed normal magnetic field and user-defined velocities. At $t=0$  a potential field extrapolation is used to fill the coronal volume, and after each update of the normal field and velocities on the boundary, the system is allowed to relax to a new equilibrium. The boundary velocities are derived from  either (a) local correlation tracking  methods
\citep{Bingert2011,Bourdin2013,Galsgaard2015}, (b) 
 imposed density perturbations proportional to the horizontal magnetic field magnitude at the surface \citep{Jiang2011,Jiang2012}, or (c) by the method of MHD characteristics \citep{Wu2006}.
 
While these data-driven models of the solar atmosphere represent an attractive approach to determining the magnetic field in the corona, there are many limitations that need to be evaluated: The lower boundary is assumed to be zero-$\beta$ or low-$\beta$ either by reducing the driving magnetic field strength, using coronal densities and temperatures, or both, and in general, only the normal component of the magnetic field vector is used. Furthermore, these methods use the photospheric magnetic field, with the other MHD quantities being inferred using various assumptions, to calculate a series of MHD equilibria, rather than to dynamically drive an MHD simulation at the photospheric boundary. 

In this work it will be shown that this latter approach, that of driving a dynamic MHD model above the photosphere using observed MHD variables, is achievable given enough photospheric data. Furthermore, a framework is developed to test and validate such a method. 
There are multiple sources of error associated with the estimation of the coronal magnetic field using this data-driving approach. Two such sources are: (1) 
inferring the photospheric plasma velocity, density and temperature when one only has measurements of the magnetic field at the photosphere, and (2) limited cadence of the photospheric data. This study  focuses on the latter of these two issues here: Suppose one has exact measurements of all the necessary MHD variables to use as boundary conditions to drive a dynamic MHD model of the photosphere, chromosphere and corona, at a particular temporal frequency. How accurate can such a data-driven solution of the coronal magnetic field be, and how does the accuracy depend on the particular frequency of the driving data and on the temporal evolution of the active region's surface magnetic field? These are important questions, as the answers can tell us whether the standard HMI time interval of 12 minutes is sufficient to accurately describe the coronal magnetic field above active regions using data-driven MHD modeling. It can also guide us as to which active regions are appropriate to study with such methods, and which evolve too quickly to be accurately modeled at such a driving interval.

Ideally, one would test the accuracy of such a data-driven simulation by comparing the estimated MHD solution with the actual  temporal evolution of the MHD variables in the domain above the surface. Obviously, this latter ground-truth dataset does not exist for solar observations. 
In this study, in lieu of this ground-truth, a dynamic MHD simulation of active region formation is run. This domain extends from below the surface into the corona, and models the emergence of magnetic flux from the upper layers of the convection zone through the stratified layers of the photosphere and chromosphere, into the corona.  Time-series of photospheric data are extracted from this \textit{``ground-truth''} dataset. These \textit{``pseudo-magnetograms"} (which also include the velocity, density, and temperature) are used as a dynamic lower boundary condition for a new \textit{``data-driven"} run which extends from the photosphere to the corona. 

\begin{figure}
\begin{center}
\includegraphics[width=3.5in]{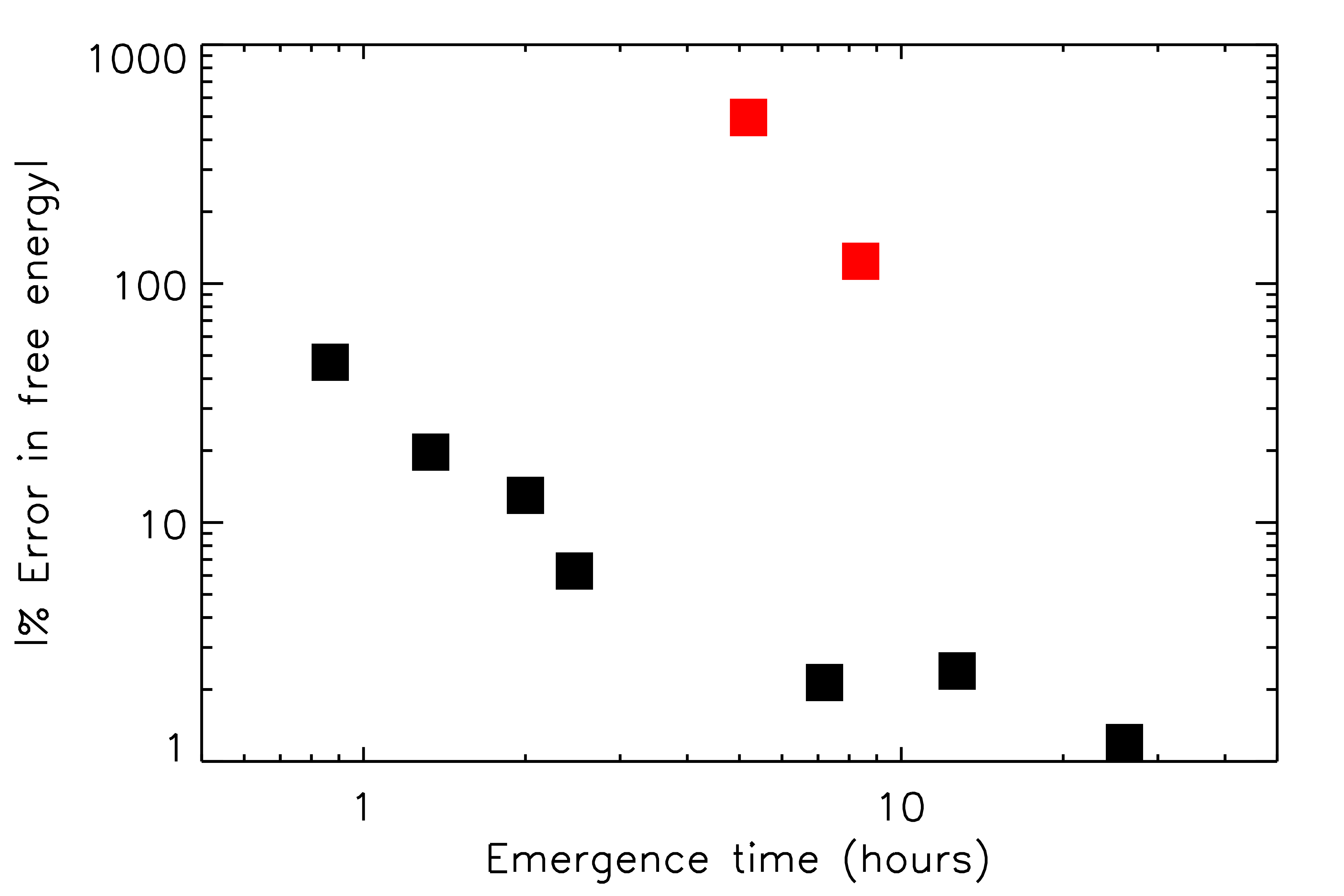}
\vskip-15pt
\caption{Absolute value of the error in the free magnetic energy in a data-driven run of simulated active region emergence, compared to the ground-truth run, as a function of the time it takes the active region to emerge. The data-driven runs use a driving interval of 12 minutes, the standard temporal resolution of available HMI magnetograms.
\label{fig:NEW_errors_time}}
\end{center}
\end{figure}

Using this framework the accuracy of the data-driving method is calculated, by comparing the ground-truth and data-driven runs, using various error diagnostics, and inferences are made on the required temporal information required to obtain satisfactory results with real observational data.

Figure \ref{fig:NEW_errors_time} shows the main results from our investigation, and plots the absolute value of the percentage error in the free magnetic energy, comparing the data-driven MHD solution to the ground-truth dataset, for a data-driving interval of 12 minutes, and for varying emergence timescale  (see \S \ref{Sim02_errors} for the definition of this error measure). The range of emergence timescales is spanned by varying the flux, twist, and initial depth of the convection zone magnetic field in the ground-truth run. The figure highlights that, generally, the longer that an active region takes to emerge, the more accurate the data-driven MHD solution can be, when using a fixed driving interval of 12 minutes. 

For active regions that emerge over long timescales the error in the free magnetic energy can be as low as 1\%.
The two red points are exceptions to this rule and, as will be shown, are results from active regions where small scale features vary on timescales shorter than 12 minutes. The details of the procedures used to generate this figure are contained in the following sections, in which the methodology of using self-consistent  dynamic MHD simulations of AR formation to test data-driven coronal models is prescribed, and results are presented on how the driving data interval affects the accuracy of the data-driven MHD solution.

\section{Using  MHD Simulations to Test the Accuracy of Data-Driven Models of the Magnetic Field Above Active Regions}

\subsection{Ground Truth MHD Simulations}
\label{ground-truth}

\begin{figure}
\begin{center}
\includegraphics[width=0.49\textwidth]{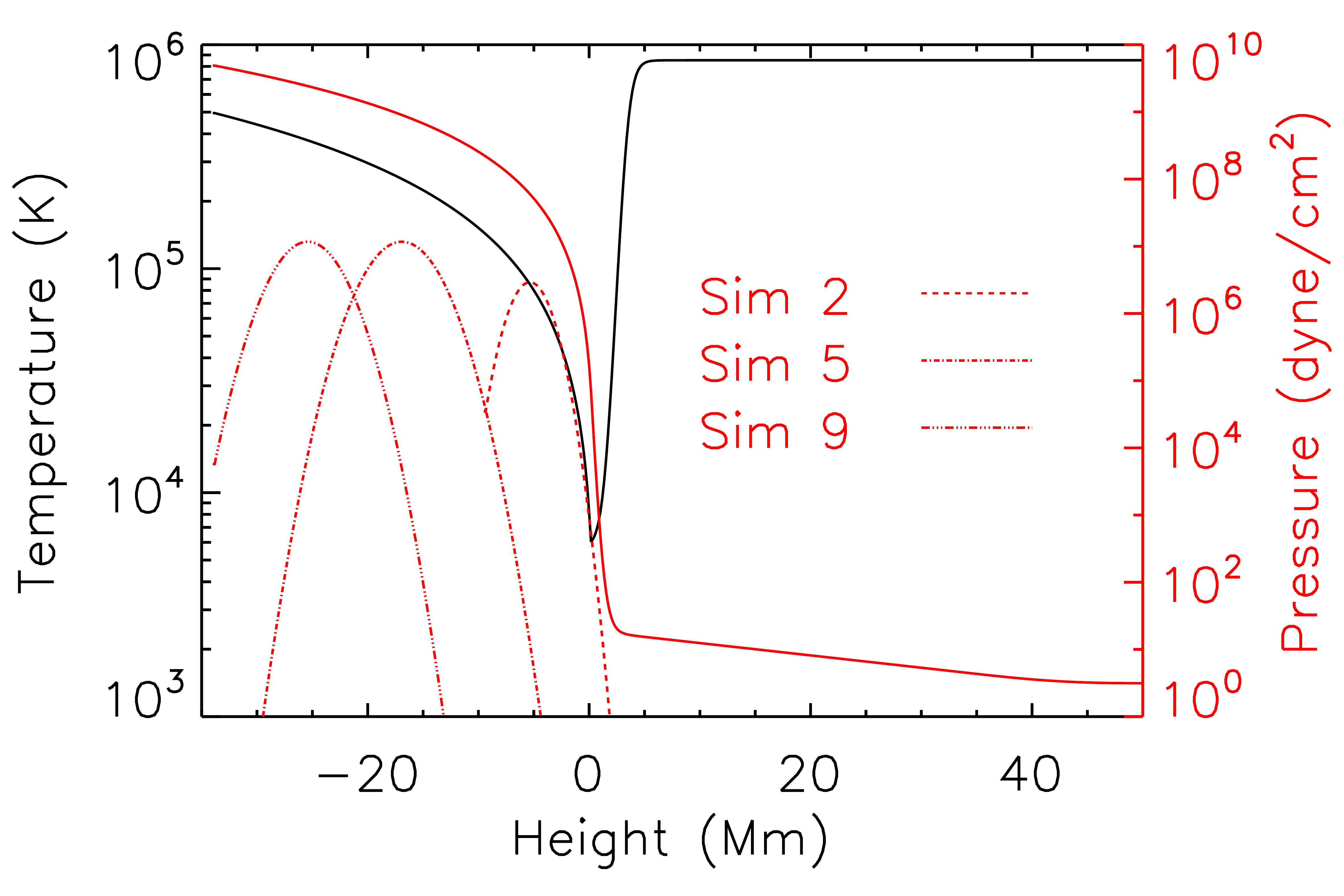}
\caption{Initial hydrostatic equilibrium. The three Gaussian-shaped  curves are the magnetic pressure of the flux tube in Simulations 2, 5, and 9.  \label{fig:IC}}
\vskip-15pt
\end{center}
\end{figure}

Our ground-truth dataset is a simulation of the dynamic emergence of a twisted flux tube from the convection zone into the corona. 
This type of simulation has been used previously to understand the emergence of magnetic flux though the convectively stable layers of the photosphere and above, the formation of bipolar active regions, the observed shearing and rotational motions seen in active regions, the build up of coronal electrical current, and the creation of stable and unstable coronal flux ropes \citep[see living review by][]{lrsp-2014-3}. The equations solved and the details of the setup of the simulation can be found in \citet{Leake_2014}, and the details of the setup are merely {\jl summarized} here. The simulations are performed using the Lagrangian Remap (LaRe) code \citet{2001JCoPh.171..151A} which evolves the MHD variables on a staggered cartesian grid. The initial {\jl thermodynamic state is a hydrostatic atmosphere consisting of a model convection zone, photosphere/chromosphere, transition region, and corona, and this is the same in all the simulations in this study.} A 1D slice of this initial atmosphere can be seen in Figure \ref{fig:IC}. Into the model convection zone a buoyant twisted magnetic flux tube structure is inserted, aligned along the $\hat{\mb{y}}$ direction:
\be
\mb{B} = B_{t}\left(-q(z-z_t)e^{-(r/a)^2},e^{-(r/a)^2},qxe^{-(r/a)^2}\right)
\ee
where $y$ is the axial direction of the tube, $z$ is the vertical direction, $B_{t}$ is the axial field strength 
at the center of the tube, $a$ is the width of the Gaussian profile, $q$ is the twist, and $z_{t}$ is the tube depth. The domain is 87Mm in the horizontal (x) direction, and extends from -10Mm to 50Mm in the vertical (z) direction for simulations 1-4 and from -37Mm to 50Mm for simulations 5-9, where z=0 denotes the photosphere.

The majority of previous flux emergence simulations, including those in \citet{Leake_2013} and \citet{Leake_2014}, are relatively small, based on the axial flux in the tube and the size of the active region, with a typical flux  $< 10^{20}$ Mx. Some larger scale simulations have recently been performed, and suggest that the mechanism for emergence is similar to that in the smaller-scale scenarios, and that the emergence rate depends on the field strength, depth, width, and twist of the initial flux tube \citep{Norton_2016}.

 \begin{figure*}
\begin{center}
\includegraphics[width=1.0\textwidth]{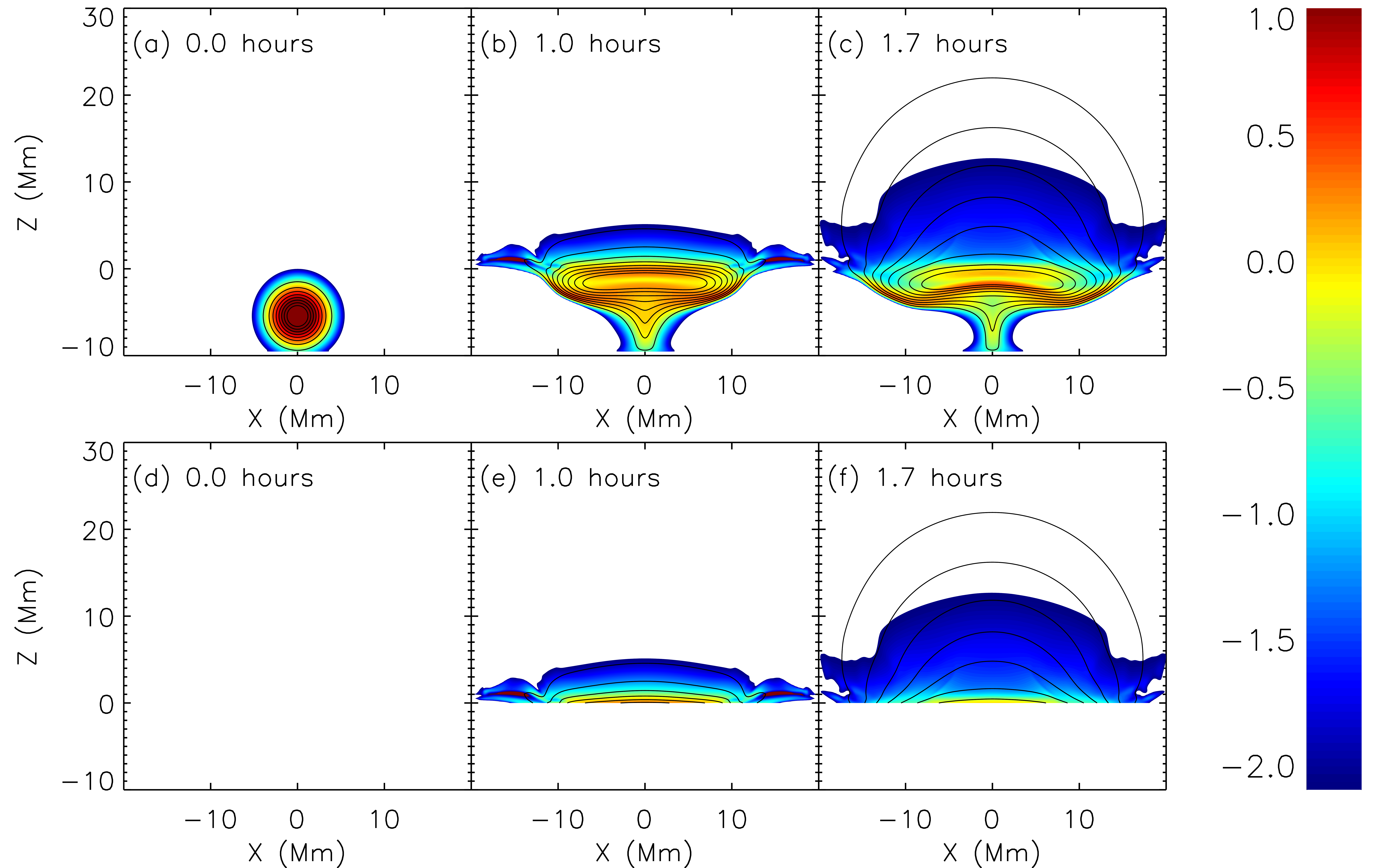}
\caption{Panels (a)-(c): Plots of $\log_{10}(B^{2}/B_{0}^2)$ where $B_{0}=1300$ G, at 0 (a), 1.0 (b), and 1.7 (c) hours for the ground-truth run for Simulation 2. Panels (d)-(f) are the same but for the data-driven run using a driving interval of 2.4 s. The black lines are contours of constant magnetic flux out of the plane. In panels (d)-(f), blank space below $z=0$ has been included, even though $z<0$ is outside the domain in the data-driven runs, as a visual aid for comparing them with the ground-truth runs.
\label{fig:2D_B}}
\end{center}
\end{figure*}

The simulations are run in 2.5D, which means there are no variations in the axial ($\hat{\mb{y}}$) direction but there is still a component of the vector fields in this direction. This allows us to run simulations for many realizations of the model parameter space and extensively investigate the accuracy of the ``data-driven'' models described below. Table \ref{table} shows the initial model parameters used for 9 ground-truth simulations, along with derived quantities in those simulations. Figure \ref{fig:IC} shows the initial magnetic pressure as a function of height for a selection of those simulations.

The details of the general evolution of this initial condition can be seen in the top panels of Figure \ref{fig:2D_B}, and are discussed in \citet{2010ApJ...722..550L}: As
 the tube rises it expands horizontally when it reaches the stable surface layers, and then emerges through these stable layers and into the corona by virtue of the magnetic buoyancy instability. 
 
\begin{figure}[ht]
\begin{center}
\includegraphics[width=0.49\textwidth]{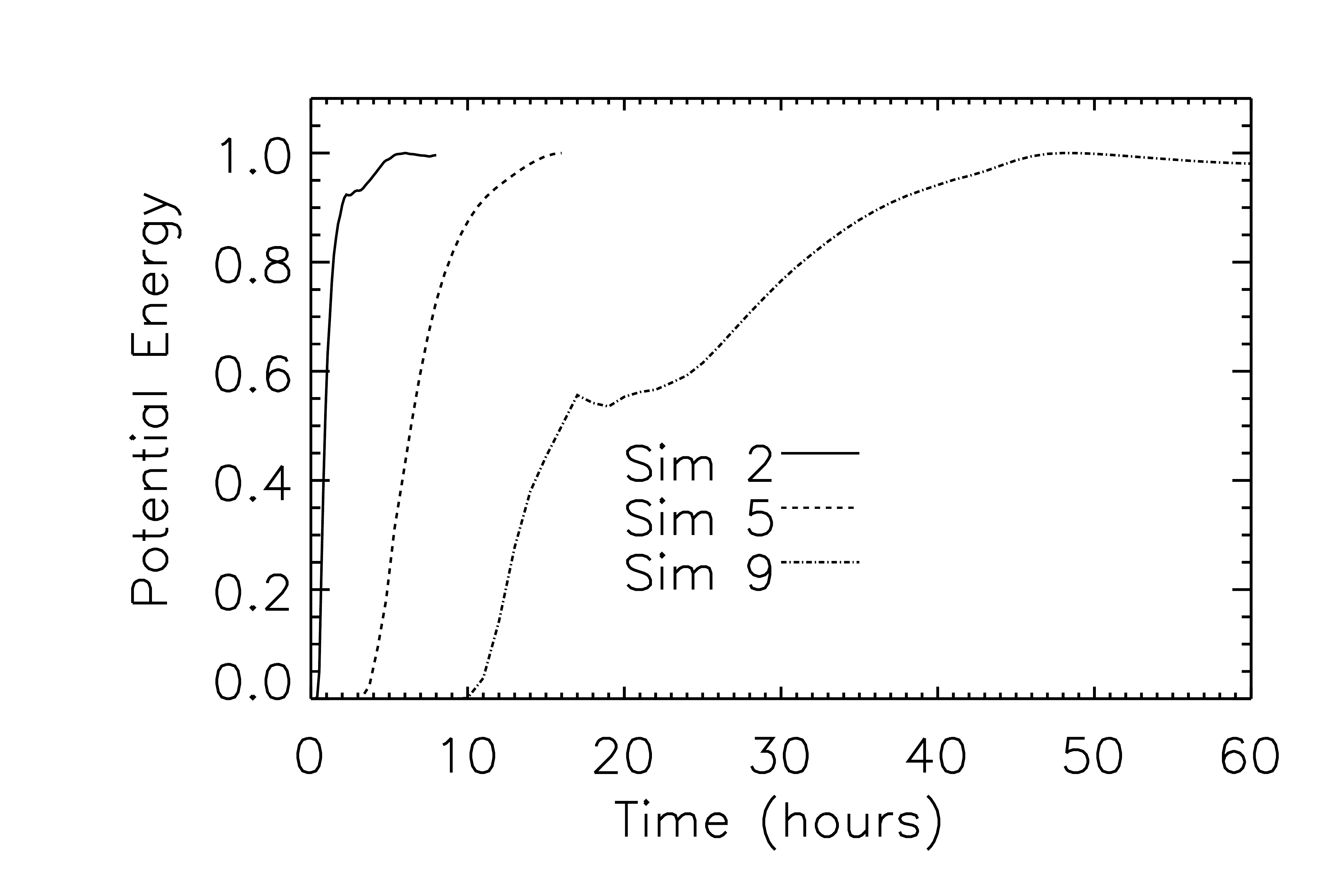}
\vskip-25pt
\caption{Potential energy above $z=0$ for Simulations 2, 5, and 9, normalized to the maximum potential energy for each simulation.  \label{fig:potE}}
\end{center}
\end{figure}

For each realization of the parameter space in Table I, the emerging active region exhibits a different emergence rate. The emergence time, $t_{emerge}$, is defined as the time elapsed between 10\% and 90\% of the maximal potential magnetic energy emerging above the surface. The potential energy above the surface is $\int_{x,z>0}{B_{pot}(x,z)dxdz}$ where $B_{pot}$ is the potential magnetic field with the same normal component on the boundary as the {\jl total} magnetic field. 
Plots of the potential energy above the surface, normalized to its maximal value, are shown for Simulations 2, 5, and 9 in Figure \ref{fig:potE}. As can be seen in Table I, our parameter study provides a range of emergence times from less than 1 hour to $\sim$ 25 hours, and so allows us to make conclusions regarding active regions which form on a wide range of timescales.

\begin{table*}[t]
\begin{center}
\begin{tabular}{ | c | c | c | c | c | c | c || c | c |}
  \hline			
  Sim. & $B$ field & Tube width & Tube twist &  Tube depth   & Beta & Emergence time & Free energy error \\
  &  $B_{t}$ (kG) &  $a$ (Mm)& qa & $z_{t}$ (Mm) &  $\beta_{tube}$ & $t_{emerge}$ (hours)  &$1-\epsilon_{free}$ (\%) \\
  \hline
  1 & 3.90 & 0.85 & 1.00 & -4.08 & 51.06 & 0.83 & +46.9 \\
  2 & 6.50 & 2.25 & 0.50 & -5.44 & 29.80 & 1.27 & +19.7 \\
  3 & 5.00 & 2.25 & 0.50 & -5.44 & 51.06 & 1.91 & +13.0 \\
  4 & 3.90 & 2.25 & 0.50 & -5.44 & 84.55 & 2.36 & +6.27 \\
  {\color{red} 5} & {\color{red} 13.0} & {\color{red} 4.25} & {\color{red} 0.50} & {\color{red} -17.0} & {\color{red} 144.0} & {\color{red} 4.98} & {\color{red} -496.0} \\
  6 & 13.0 & 4.25 & 0.50 & -20.4 & 230.1 & 6.90 & +1.50 \\
  {\color{red} 7} & {\color{red} 13.0} & {\color{red} 4.25} & {\color{red} 0.40} & {\color{red} -20.4} & {\color{red} 230.5} & {\color{red} 8.05} & {\color{red} -124.0} \\
  8 & 8.00 & 4.25 & 0.50 & -20.4 & 609.0 & 12.17 & +1.30 \\
  9 & 8.00 & 4.25 & 0.50 & -20.4 & 609.0 & 24.92 & +1.20 \\
  \hline  
\end{tabular}
\caption{Simulation parameters and results, ordered by emergence time. The red rows are particular simulations discussed later in this paper. The final column are results from analysis of the error in the associated data-driven run, discussed  in \S \ref{sec:param}. $\beta_{tube}$ is defined at the center of the tube. \label{table}}
\end{center}
\end{table*}

\subsection{Data-Driven Simulations}

The MHD simulations described in \S\ref{ground-truth} model the evolution of magnetic field
as it buoyantly rises from the upper layers of the convection zone through the stratified layers of the photosphere and chromosphere, and into the corona. Time-series of photospheric data are then  extracted from this ground-truth dataset. 
These \textit{``pseudo-magnetograms"} (which also include the velocity, density, and temperature) are used as a dynamic lower boundary condition for a new \textit{``data-driven"} run which extends from the photosphere to the corona. The ground-truth and data-driven runs are then compared above the surface to test the accuracy of the data-driving method, using certain error diagnostics.
Provided enough temporal and spatial resolution is applied, and provided the lower boundary  is driven consistently with the MHD equations, the magnetic field above the photosphere in the data-driven and ground-truth runs should be the same (within the typical accumulation of machine precision errors over space and time).

\begin{wrapfigure}[20]{r}{1.55in}
\vskip-30pt
\includegraphics[width=1.5in]{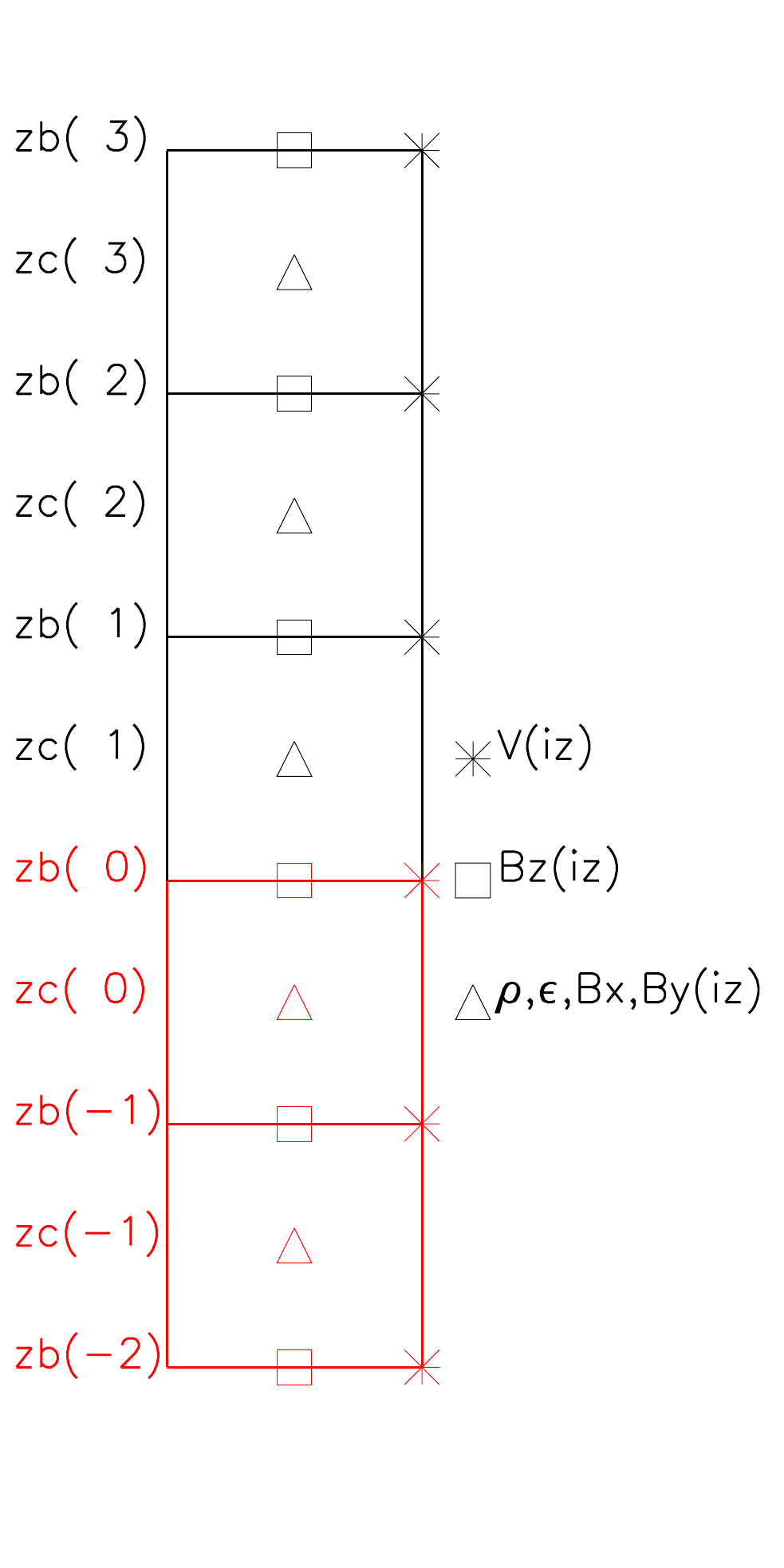}
\vskip-40pt
\caption{Vertical grid of the LaRe code. The boundary and ghost cells are colored red. \label{fig:Lare2d_boundary}}
\vskip-10pt
\hrulefill
\end{wrapfigure}

Figure \ref{fig:Lare2d_boundary} shows where the MHD variables are evolved on the staggered 2.5D LaRe grid. The cell center for the $iz$-th vertical cell is located at a height of $z_{c}(iz)$ and the cell edge is located at $z_{b}(iz)$. Density ($\rho$), energy ($\epsilon$), and the horizontal magnetic field components ($B_{x},B_{y}$) are defined at cell centers. The vertical magnetic field ($B_{z}$) is defined on the top cell edge, and the velocity vector $\mb{V}$ is defined on the cell vertex.

MHD only requires specification of variables on the lower boundary of the data driven runs at zb(0).
However, presently, LaRe treats the boundary cells of the driven simulation with the same equations as the interior cells of the simulation.
Therefore, because of the 3rd order gradients used in the remap step in LaRe, the integration method requires information
from two surrounding cells in each direction to evolve the system. For example, evolution of values at zc(1) requires information from zc(0) and zc(-1) which are formally exterior to the driven simulation. 
{\jl This approach, in a study which is focused on the effects of the temporal cadence of the driving data, allows for a proof of concept of the data-driving method and allows the specification of errors associated with insufficient temporal information. This approach, however, may over-specify the boundary information in certain conditions. Future work will investigate particular choices of boundary implementation, such as the method of characteristics \citep[e.g.,][]{Wu2006}  to determine which MHD variables are required at the boundary.}

In the data-driven run, the boundary is supplied with {\jl all the MHD variables (magnetic field, velocity, pressure, and temperature)} at discrete intervals, and between those times interpolation is used to update the boundary conditions for each iteration of the numerical code. Figure \ref{fig:2D_B} shows the result of this data-driving process at various times during the emergence of magnetic flux from beneath the surface, for Simulation 2. The snapshots are taken at 1, 1.7, and 2.4 hours. The top panels show the ground-truth emergence run, and the bottom panels show the data-driven run, when the pseudo-magnetograms are applied to the data-driven boundary every 2.4 s, which is much smaller than the emergence timescale of 1.27 hours (Table I), and is approximately equal to the initial CFL \citep{CFL_paper} limited timestep in the simulation.  {\jl This timestep is defined as $dt_{CFL} = min(dx/|\mb{V}|, dx/C_{f})$ where $C_{f} = \sqrt{C_{s}^2+C_{A}^2}$, where dx is the grid-cell resolution (170km), and where $C_{s}$ and $C_{A}$ are the local acoustic and Alfv\'{e}n speed, respectively. }

Figure \ref{fig:2D_B} shows that the data-driving process is successful on a fundamental level: the magnetic flux appears at the photosphere in the data-driven run where no significant flux was initially present at all in the domain. Not only that, but the flux distribution it generates within the domain looks visually to be the same as what is generated in the ground truth run, where it is self-consistently generated by flux emergence, rather than by boundary driving. 

In the next section, the simulations will be examined using quantitative error analysis, focusing on how the errors vary when one drives at slower cadences, which are more representative of cadences associated with standard photospheric observations.

\begin{figure*}
\begin{center}
\vspace{-5mm}
\includegraphics[width=\textwidth]{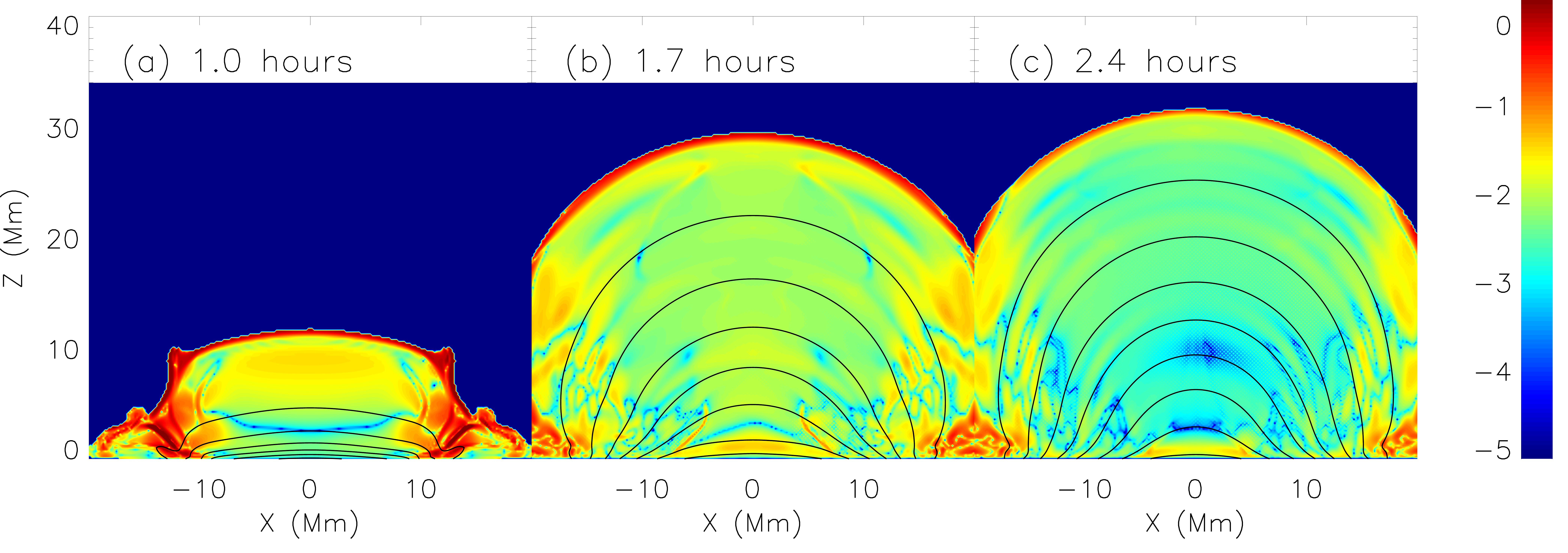}
\caption{Spatial representation of the error in the data-driven run, given by  $\log_{10}{(|1-\mb{b}_{i}^{2}/\mb{B}_{i}^{2}|)}$, for Simulation 2 with a driving interval of 0.04 minutes, or 2.4 seconds. \label{fig:2D_error}}
\end{center}
\end{figure*}

\section{Results}

In this section results of quantitative error analysis are presented, first focusing in detail on one simulation (Simulation 2) that has an emergence timescale of 1.27 hours, and then expanding to the nine simulations listed in Table I. For each Simulation,  the frequency at which the pseudo-magnetograms from the ground-truth run are used to drive data-driven runs is varied.

\subsection{Case-Study of Simulation 2}
\label{Sim02_errors}

For a better visual comparison of the 2.4s data-driven and the ground-truth run for Simulation 2, Figure \ref{fig:2D_error} shows the difference in the two magnetic field solutions at the same three moments in time (1, 1.7, and 2.4 hours).  
The difference is calculated at each pixel ($i$) in the 2D driven domain as $\log_{10}{(|1-\mb{b}_{i}^{2}/\mb{B}_{i}^{2}|)}$ where $\mb{B}_{i}$ is the ground-truth magnetic field and  $\mb{b}_{i}$ is the data-driven magnetic field. The errors appear maximal at the edge of the emerging structure, with lower errors inside the emerging magnetic field.

To further compare the datasets of magnetic field between the ground-truth and data-driven runs, 
a set of error metrics based on some of the scalar functions used in \citet{2006SoPh..235..161S} is used:
\begin{eqnarray}
1-C_{vec}&  = &  1 - \sum_{i}(\mb{b}_{i}\cdot\mb{B}_{i}) \Big/  \sqrt{\sum_{i}{{b}_{i}^2}}\sqrt{\sum_{j}{B_{j}^2}} \nonumber \\
E_{n} & = & \sum_{i}|\mb{b}_{i}-\mb{B}_{i}| \Big/ \sum_{i}|\mb{B}_{i}| \nonumber \\
1-\eps &  = & 1- \sum_{i}|\mb{b}_{i}|^{2} \Big/ \sum_{i}|\mb{B}_{i}|^{2}  \nonumber \\
1-\phi_{y} & = & 1 - \sum_{i}{b}_{y,i} \Big/ \sum_{i}{B}_{y,i}\nonumber  \\
1-\phi_{x} & = &  1 - \sum_{i,x=0}{b}_{x,i}\Big/ \sum_{i,x=0}{B}_{x,i} \nonumber \\
1-\eps_{free} & = & 1 - \sum_{i}\left(|\mb{b}_{i}|^{2}-|\mb{b_{pot}}_{i}|^{2}\right) \Big/ \nonumber \\
& &  \sum_{i}\left(|\mb{B}_{i}|^{2}-|\mb{B_{pot}}_{i}|^{2}\right)
 \label{eqn:metrics}
\end{eqnarray}
where the subscript $pot$ represents the potential magnetic field. These functions represent, respectively, the vector correlation ($1-C_{vec}$), the normalized vector error ($E_{n}$), the error in the total magnetic energy ($1-\epsilon$), the errors in the out-of-plane ($y$) and horizontal ($x$) flux ($1-\phi_{y}, 1-\phi_{x}$), and the error in the free magnetic energy ($1-\epsilon_{free}$). The last metric is a new metric, equal to 1 minus the ratio of free magnetic energy in the data-driven run to that in the ground-truth run. Note that if $\mb{b}_{i}=\mb{B}_{i}$ everywhere,  then all of these functions are identically zero.

\begin{figure}[h]
\begin{center}
\includegraphics[width=0.49\textwidth]{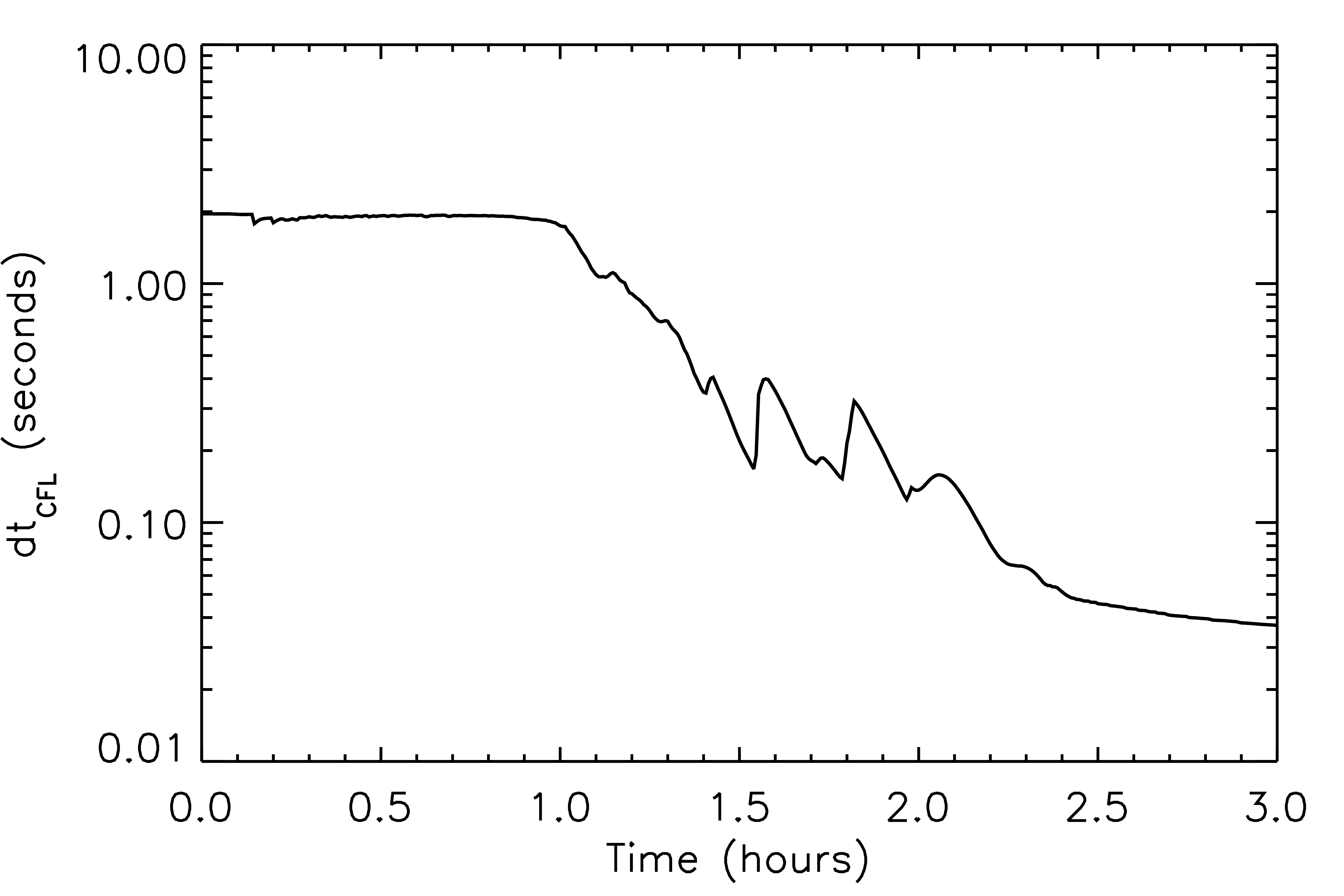} 
\caption{CFL limited timestep as a function of time for Simulation 2. \label{fig:CFL}}
\end{center}
\end{figure}

To investigate the effect of the temporal frequency of the driving data on the accuracy of the data-driven solution, the frequency with which boundary data is supplied to the data-driven runs is varied, and linear interpolation is used at steps between data inputs. The data-driven and ground-truth runs are compared above the surface to test the accuracy of the data-driving. Primary driving timescales of [0.04, 0.12, 0.4, 1.2, 4, 12] minutes are used, the largest being the standard timescale of the HMI magnetic field data. Two additional data-driven runs are performed. The first is driven every $dt_{CFL}$ where $dt_{CFL}$ is the CFL limited timestep of the numerical scheme.  Figure \ref{fig:CFL} shows $dt_{CFL}$ as a function of time for Simulation 2. The timestep decreases significantly when magnetic field emerges into the low density corona and the Alfv\'{e}n speed increases. The second additional data-driven run is provided with boundary data at every substep within one CFL iteration of the numerical code where an interior MHD {\jl variable} is updated. This particular \textit{``sub-CFL''} data-driven run should thus result in the same atmosphere above $z=0$, with errors equal to spatially and temporally accumulated machine precision errors for the chosen error metric.

\begin{figure}[h]
\begin{center}
\includegraphics[width=0.49\textwidth]{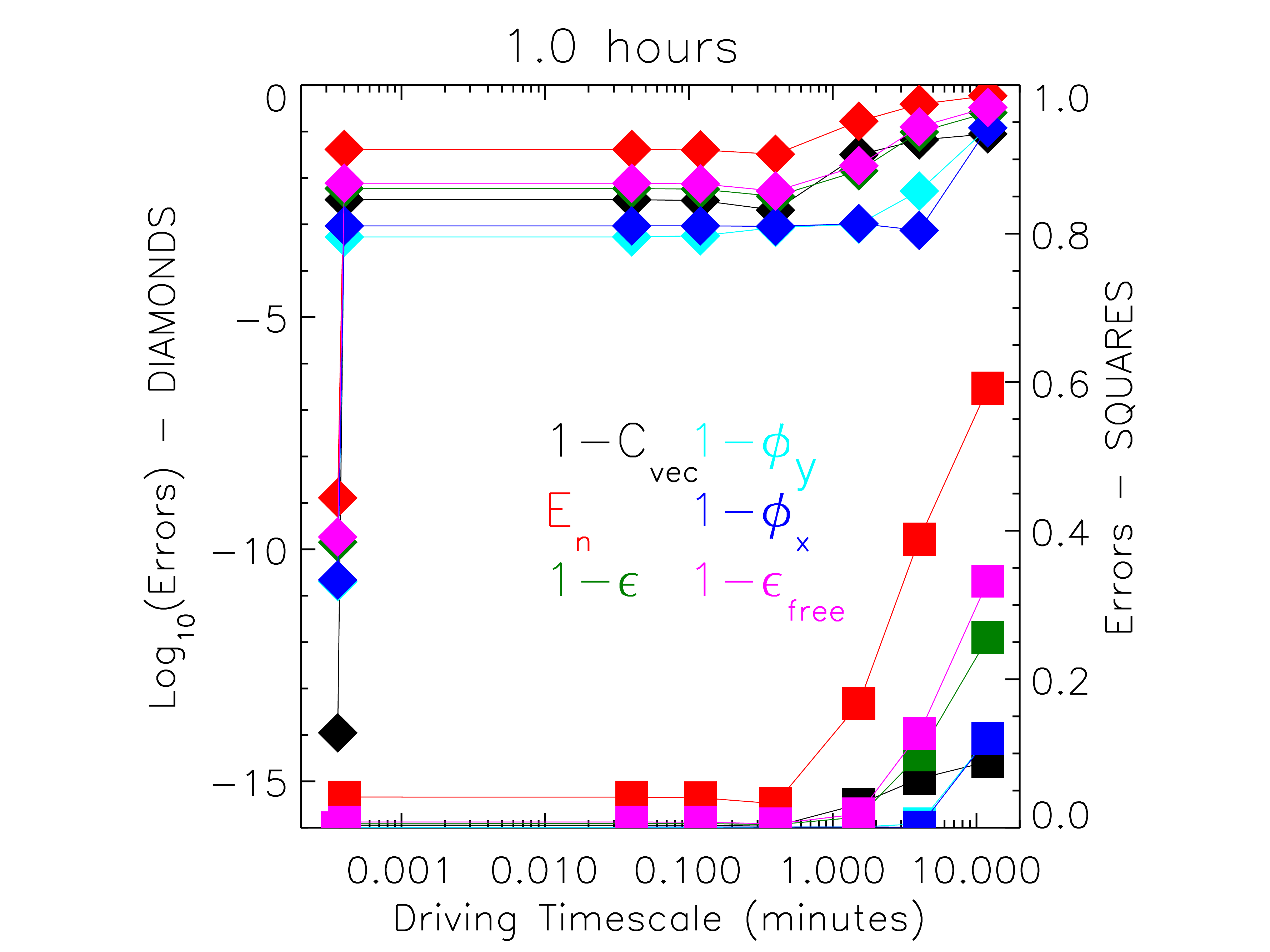} 
\caption{Magnitude of the error metrics for the data-driven runs, at 1 hour into the run, as a function of the driving interval, for the case of Simulation 2. Diamonds are on a log scale and the squares are on a linear scale. The data points at 0.003 and 0.004 minutes are for the sub-CFL and CFL timesteps, respectively. \label{fig:errors_sims1}}
\end{center}
\end{figure}

\begin{figure}[h]
\begin{center}
\includegraphics[width=0.49\textwidth]{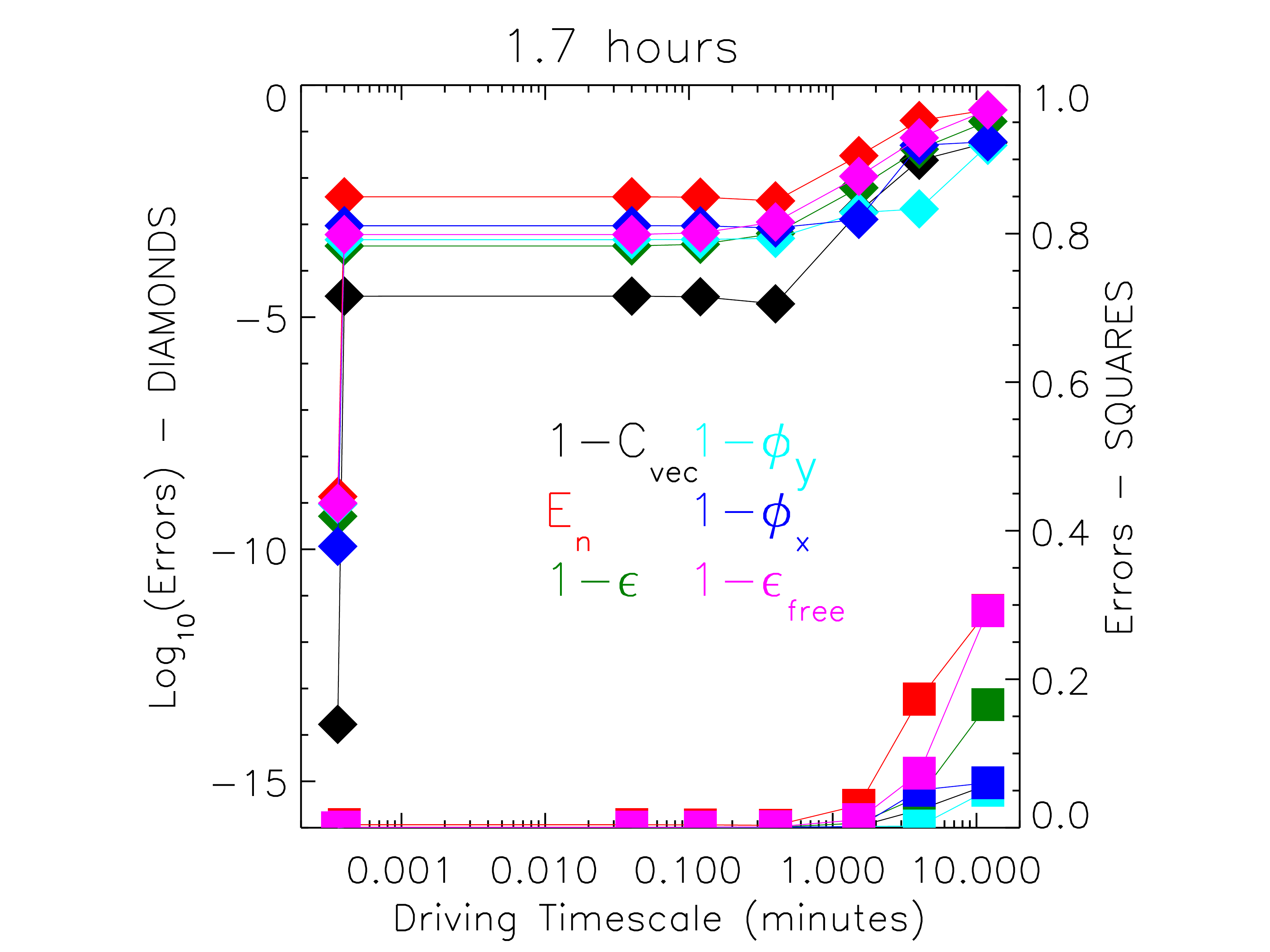} 
\caption{Same as Figure \ref{fig:errors_sims1} but for a time of 1.7 hours. \label{fig:errors_sims2}}
\end{center}
\end{figure}

\begin{figure}[h]
\begin{center}
\includegraphics[width=0.49\textwidth]{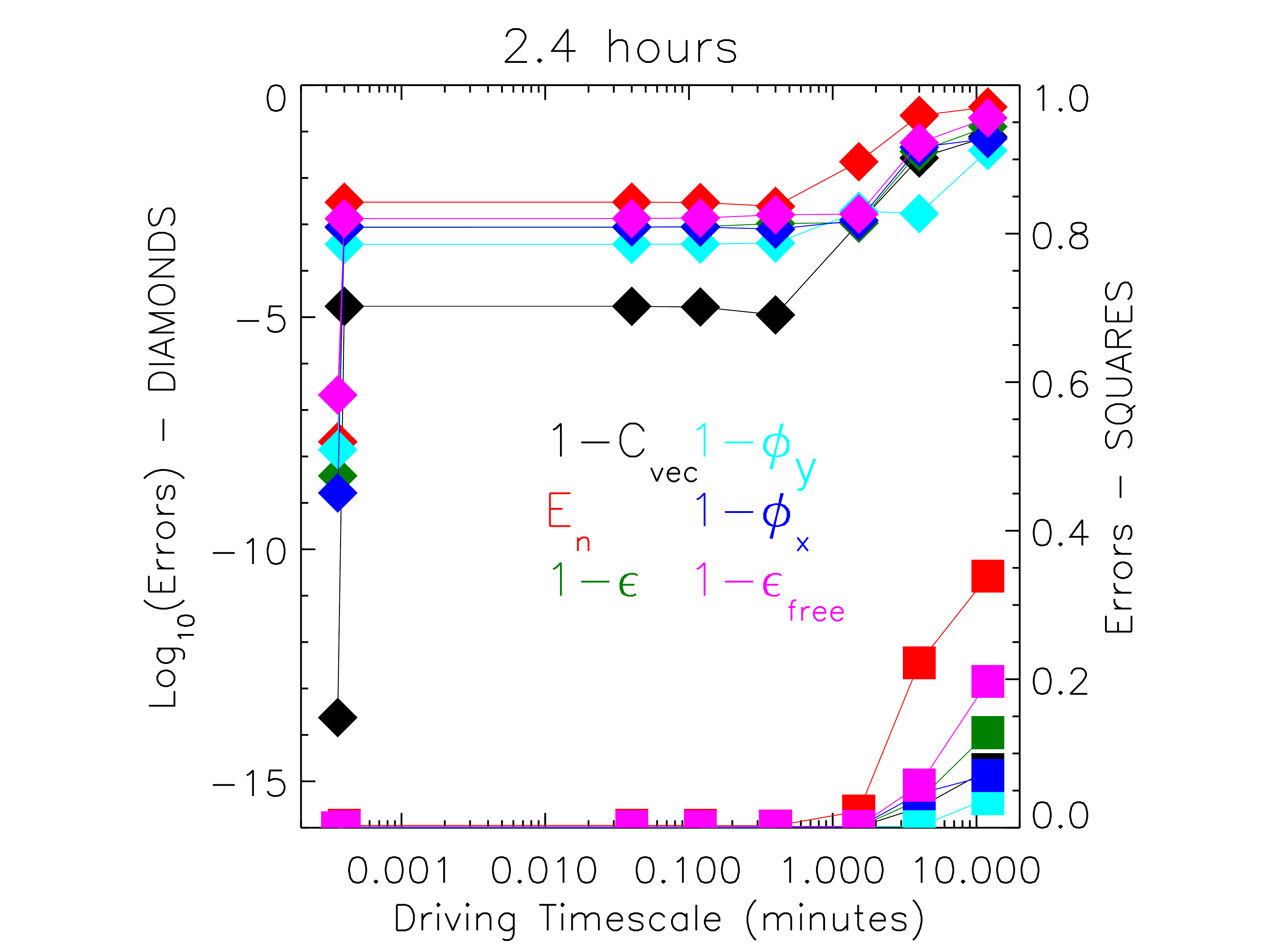} 
\caption{Same as Figure \ref{fig:errors_sims1} but for a time of 2.4 hours. \label{fig:errors_sims3}}
\end{center}
\end{figure}

Figures \ref{fig:errors_sims1}-\ref{fig:errors_sims3} show the magnitude of the metrics of Equations (\ref{eqn:metrics}) for Simulation 2, at 1, 1.7, and 2.4 hours into the emergence, as a function of the driving interval. The errors are shown on logarithmic (left axis, diamonds) and linear (right axis, squares) scales. The two additional data-driven runs, in which the boundary is driven with data on the sub-CFL and CFL limited timestep have driving timesteps which vary throughout the simulation, but are plotted for illustrative purposes at times of .0003 and .0004 minutes, respectively.

Using the the linear scale in Figures \ref{fig:errors_sims1}-\ref{fig:errors_sims3},  one can see that the errors are insignificant at driving intervals less than 1 minute, which is much less than the emergence timescale of 1.27 hours for Simulation 2 (Table I). Above a driving interval of 1 minute,  the errors increase, with, e.g., the error in the free energy (magenta line) 
reaching 30\% at 1.7 hours into the emergence, when driven at a 12 minute cadence (Figure \ref{fig:errors_sims2}).

Using the logarithmic scale in Figures \ref{fig:errors_sims1}-\ref{fig:errors_sims3}, and focusing on the metric $1-C_{vec}$ at 2.4 hours (black lines of Figure \ref{fig:errors_sims3}), one can see that with decreasing driving interval, the magnitude of the error plateaus at about $10^{-5}$ for Simulation 2, even at the CFL limited timestep (represented by the datapoint at .0004 minutes).  For the sub-CFL driving interval (represented by the datapoint at .0003 minutes) however, the error drops to $10^{-14}$.  The other metrics show larger magnitudes at this sub-CFL interval, as their integrals are cumulative in nature. This analysis shows that the method of data-driving, when using all the available  boundary data as needed by the numerical algorithm,  is able to reproduce the ground-truth solution to within errors associated with accumulation of machine precision, and allows confident conclusions to be drawn at longer driving intervals.
 
The free magnetic energy is an important measure for space weather, as it gives an approximate measure for the amount of magnetic energy that can be released in a solar flare, CME, or filament eruption, and so warrants error analysis in these data-driven runs. However, the free magnetic energy is defined as the difference between the total magnetic energy and the energy in the potential magnetic field, where the potential field is the field having the same vertical field component at the surface as the reference field, but with no electric currents ($\mb{J}\sim\nabla\times\mb{B}=0$). Before making conclusions regarding the accuracy of the data-driven runs based on the error in free magnetic energy, it must be established that these errors are not due to errors in the potential magnetic energy. 

At specific times (e.g., every 12 minutes) when surface boundary data {\jl are} supplied from the ground-truth run to the data-driven run, the potential field in the data-driven run should be identical to the potential field above the surface in the ground-truth run. In this case the error in the free magnetic energy will be equal to the error in the total magnetic energy  multiplied by the ratio of total to free energy. This is shown below, denoting $^{dd}$ for the data-driven solution, and $^{gt}$ for the ground truth solution, and using the assumption that $E^{dd}_{pot} = E^{gt}_{pot}\equiv E_{pot}$:
\bea
1-\eps_{free} & = & 1 - \frac{E^{dd}_{free}}{E^{gt}_{free}} = 
\frac{E^{gt}_{free}-E^{dd}_{free}}{E^{gt}_{free}} \nonumber \\
& = & \frac{E^{gt}-E_{pot}-E^{dd}+E_{pot}}{E^{gt}_{free}} \nonumber \\
& = & \frac{E^{gt}}{E^{gt}_{free}}(1-\frac{E^{dd}}{E^{gt}})=
\frac{E^{gt}}{E^{gt}_{free}}(1-\eps).
\label{eq:error}
\eea

  \begin{figure}[h]
\begin{center}
\includegraphics[width=0.49\textwidth]{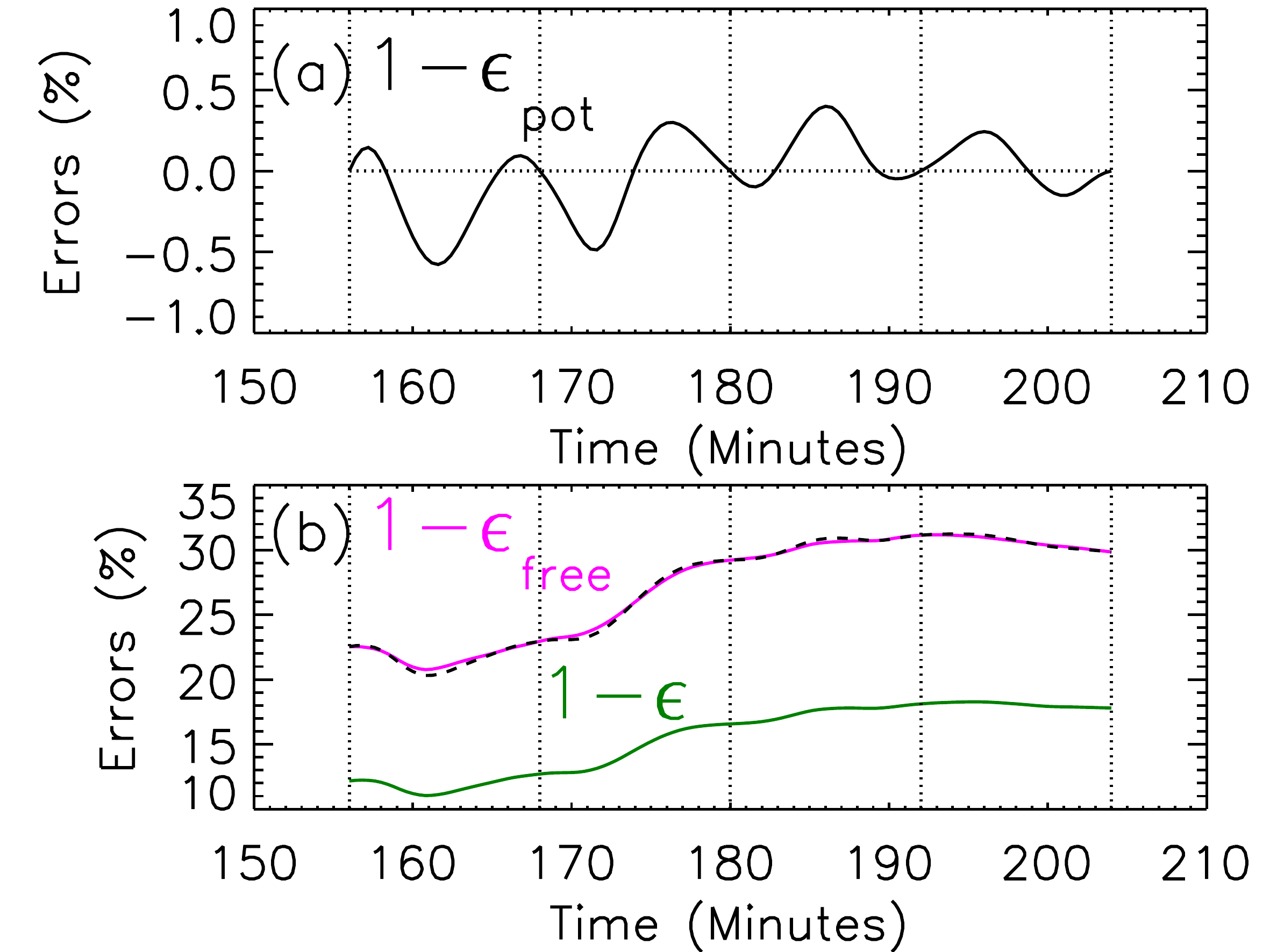}
\caption{Errors (\%) in potential (a), total and free (b) energies around the 3 hour mark for Simulation 2, when the data-driven run is provided with data every 12 minutes (indicated by vertical dotted lines). \label{fig:errors_pot_free}}
\end{center}
\end{figure}

At times between the regular data-driving inputs, errors will grow in the potential energy, as the boundary data {\jl are} not exactly equal to the ground truth data at those times, but linearly interpolated. It is necessary to check that these errors in the vertical magnetic field do not result in significant errors in the potential field.  Figure \ref{fig:errors_pot_free} shows the error in potential, total, and free energy in the data-driven run around the 3 hour mark for Simulation 2, when using driving data every 12 minutes. As predicted, the error in the potential energy is zero at data-input times, and maximal at points between these times. At these times, the absolute error in the potential energy is below 0.5\%. Noting that the errors in the free energy are approximately 20-30\% during this period, it is clear that the errors in potential energy associated with the linear interpolation between driving time inputs do not significantly contribute to the errors in the free energy. Furthermore, the dashed line in Figure \ref{fig:errors_pot_free}(b) is the total energy error multiplied by the ratio of total to free energy, as in Equation (\ref{eq:error}), and only deviates slightly from the free energy error at midpoints between data-driving inputs.  This analysis shows that calculated errors in free magnetic energy  can be reliably used to make conclusions about the accuracy of the data-driven runs in the following section.

\subsection{Parameter study}
\label{sec:param}

\begin{figure}[h]
\begin{center}
\includegraphics[width=0.49\textwidth]{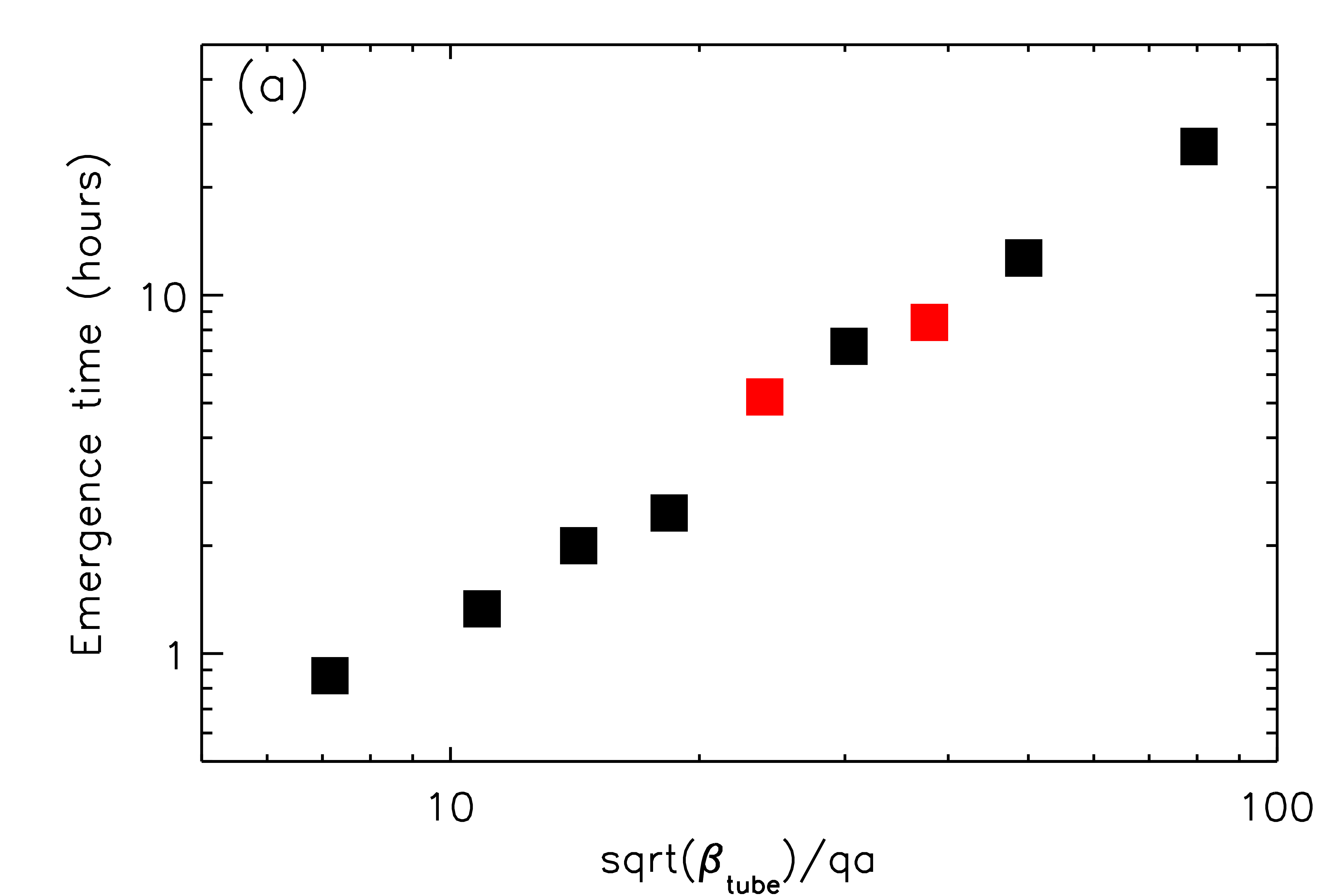}
\includegraphics[width=0.49\textwidth]{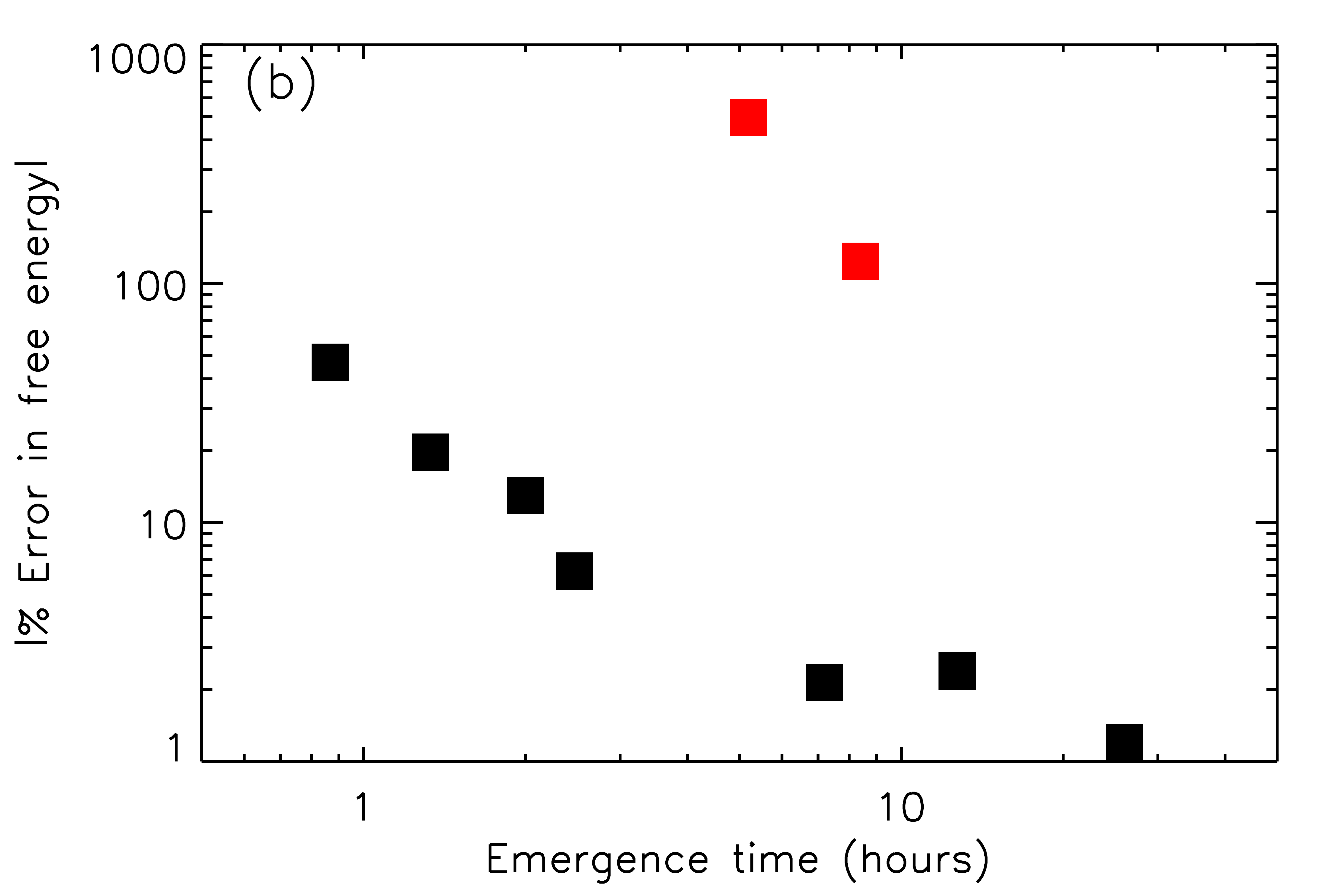}
\vspace{-10mm}
\caption{Panel (a): Emergence timescale as a function of parameters: $\sqrt{\beta_{tube}}/qa$. Panel (b): Absolute value of error in free magnetic energy versus emergence timescale.  The two red points in both panels are for Simulations 5 and 7, and exhibit large negative values of the free energy error. \label{fig:errors2}}
\end{center}
\end{figure}

 Figure \ref{fig:errors2}(a) shows the emergence time for each of the nine ground-truth simulations as a specific function of the initial condition model parameters, $\sqrt{\beta_{tube}}/qa$. 
This function is chosen based on insight from previous simulations of flux emergence in the convection zone and atmosphere above: 
\citet{2006A&A...460..909M} showed that as flux tubes rise in the convection zone, they fragment based on the amount of twist ($qa$ in these simulations) in the tube. The lower the twist, the more fragmented and distorted the tube becomes, and as a result, the  less magnetic flux is able to build up at the top of the convection zone to drive the emergence. In addition, \citet{Longcope1996} and \citet{1998ApJ...492..804E}  showed that the initial rise speed of the tube increases in proportion to  $1/\sqrt{\beta_{tube}}$, and although there is no guarantee that the rise speed influences the emergence speed, this is a reasonable function to use to predict emergence rates. Relating these insights to the initial parameters of the ground-truth runs, and based on Figure \ref{fig:errors2}(a), a monotonic dependence of the emergence time on $\sqrt{\beta_{tube}}/qa$ is suggested here, leaving 
a more complete parametric study of flux emergence to future publication.

Figure \ref{fig:errors2}(b) shows the magnitude of the error in the free magnetic energy in the data driven runs for each of the nine simulations from Table I, as a function of the emergence time, when the data-driven run is supplied with MHD data every 12 minutes. The error calculation is performed towards the end of each emergence event, when the potential energy has reached its maximal value. 
From Figure \ref{fig:errors2}(b), a general inverse relationship between the emergence timescale and the error in the free energy can be seen, with the slowest emerging Simulation (9) having a free energy error in the data-driven solution just above 1\%. This result is encouraging for the validity of this method of data-driving to model the coronal magnetic field above emerging active regions using HMI magnetograms, particularly as future data may  be available at even smaller intervals \citep{Xudong}.

However, from Figure \ref{fig:errors2}(b), it is also clear that there are two exceptions in the parameter space, Simulations 5 and 7, which show very large error magnitudes ($> 100\%$), and whose data points are colored red in both panels. As seen in Table I, the error in the free magnetic energy $(1-\epsilon_{free})$ of the data-driven run for these two simulations is both large and negative, indicating that there is much more free magnetic energy in the data-driven run than in the ground-truth run. Note also from Table I that these two simulations only differ by relatively small values of the initial flux tube parameters (and hence emergence time) from other simulations which show small ($<20\%$), positive errors in the free magnetic energy. To understand why and when such large errors might occur in data-driven models using observed magnetograms,  it is instructive to examine these two simulations further.

\begin{figure}[h]
\begin{center}
\includegraphics[width=0.49\textwidth]{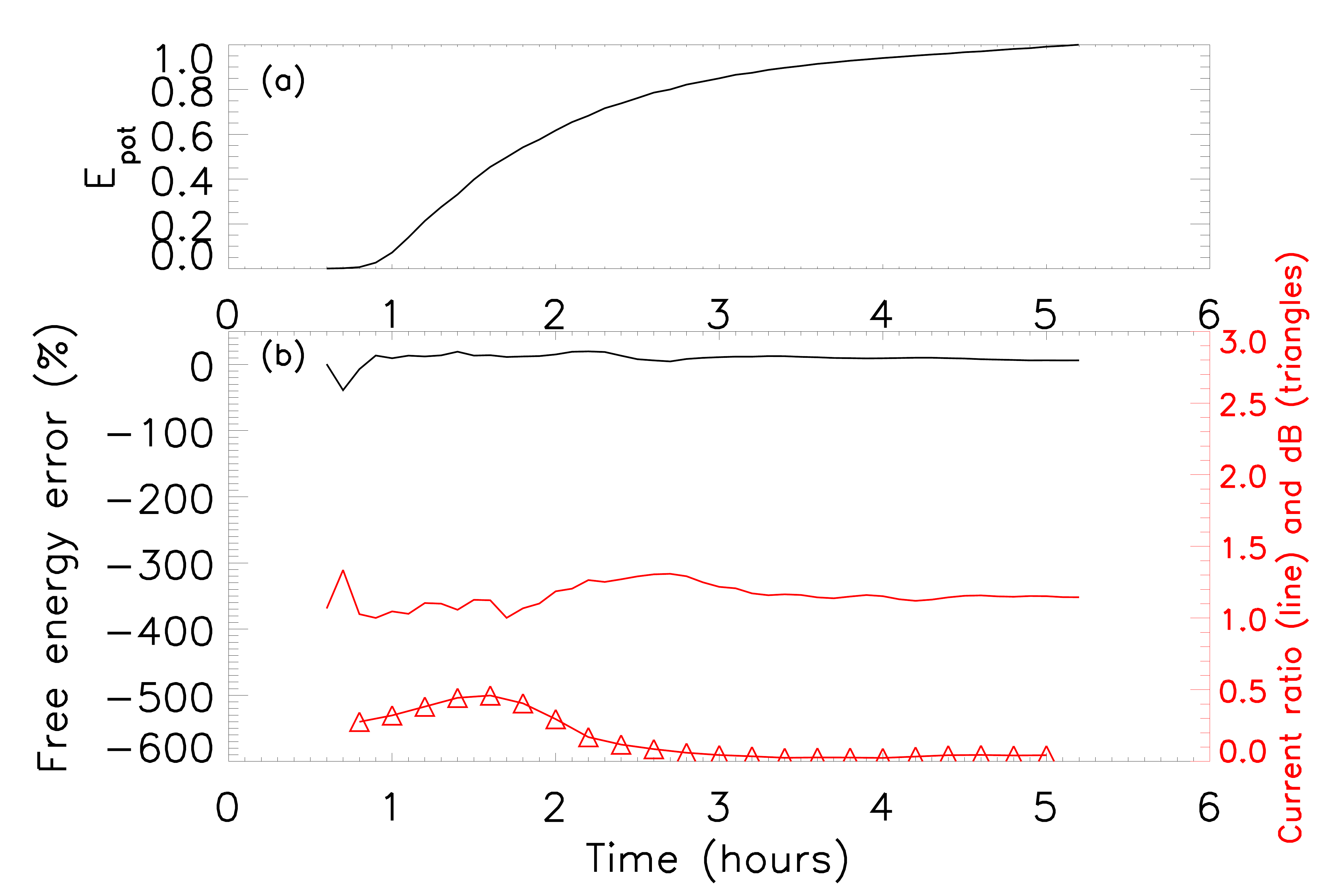}
\vspace{-10mm}
\caption{Panel (a): Evolution of potential energy in ground-truth run for Simulation 4. Panel (b): Measure of magnetic field variability $dB$ (red triangles) for the same Simulation. Error in free energy for the data-driven run (solid black line). Ratio of out-of plane current between the data-driven and ground-truth run (solid red line). \label{fig:currents1}}
\end{center}
\end{figure}

\begin{figure}[h]
\begin{center}
\includegraphics[width=0.49\textwidth]{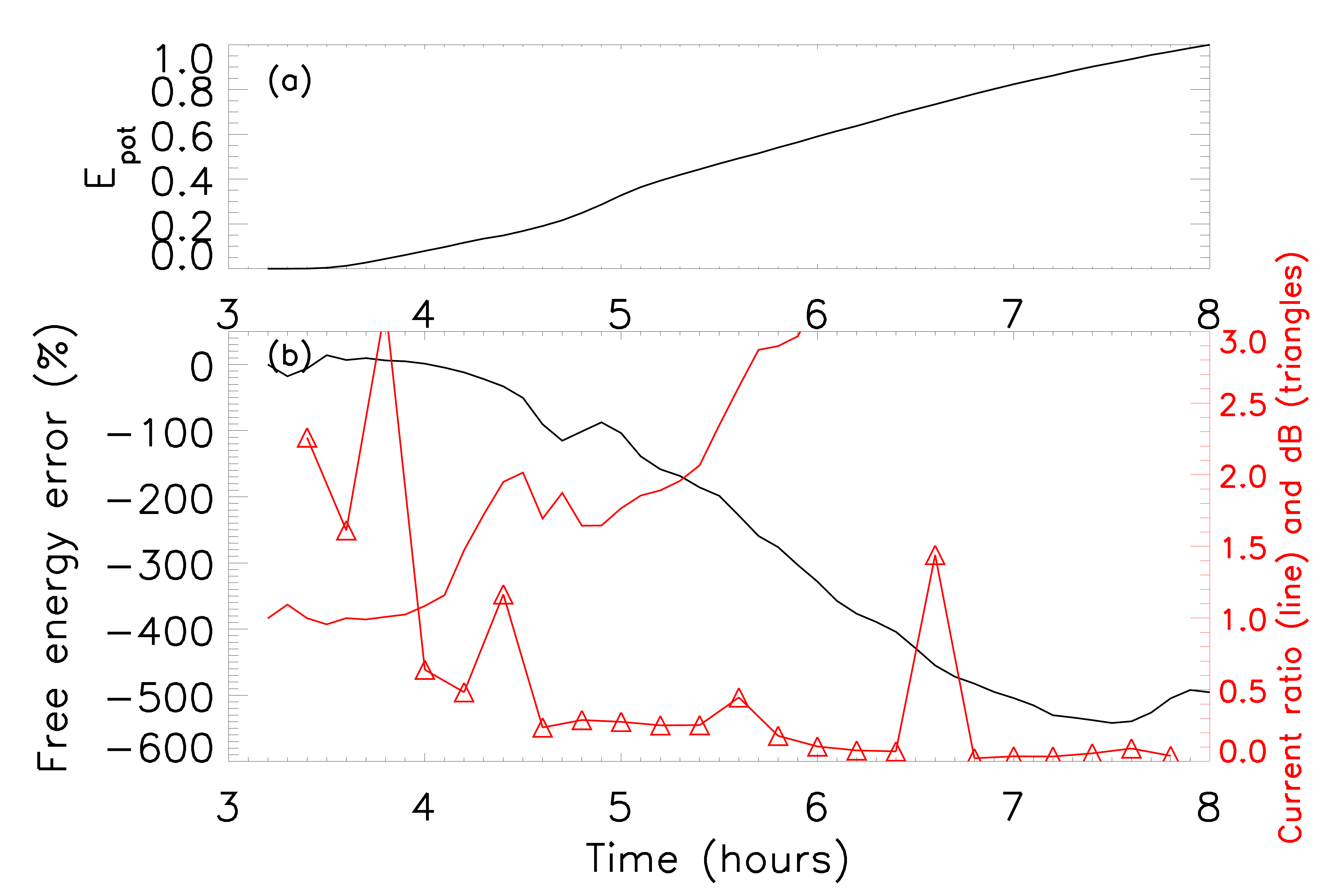}
\vspace{-10mm}
\caption{Same as Figure \ref{fig:currents1} but for Simulation 5.
\label{fig:currents2}}
\end{center}
\end{figure}

\begin{figure*}[t]
\begin{center}
\includegraphics[width=0.32\textwidth]{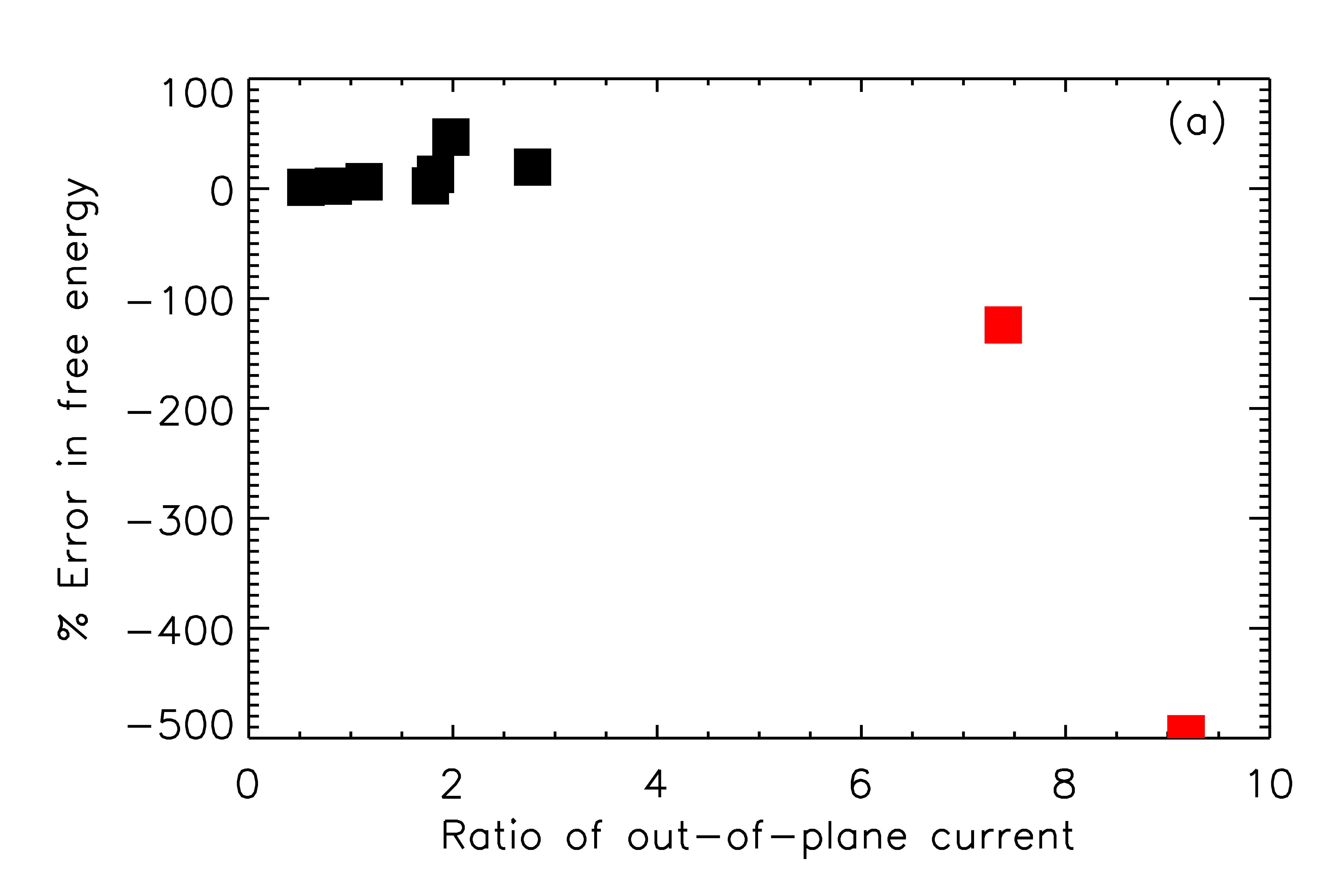}
\includegraphics[width=0.32\textwidth]{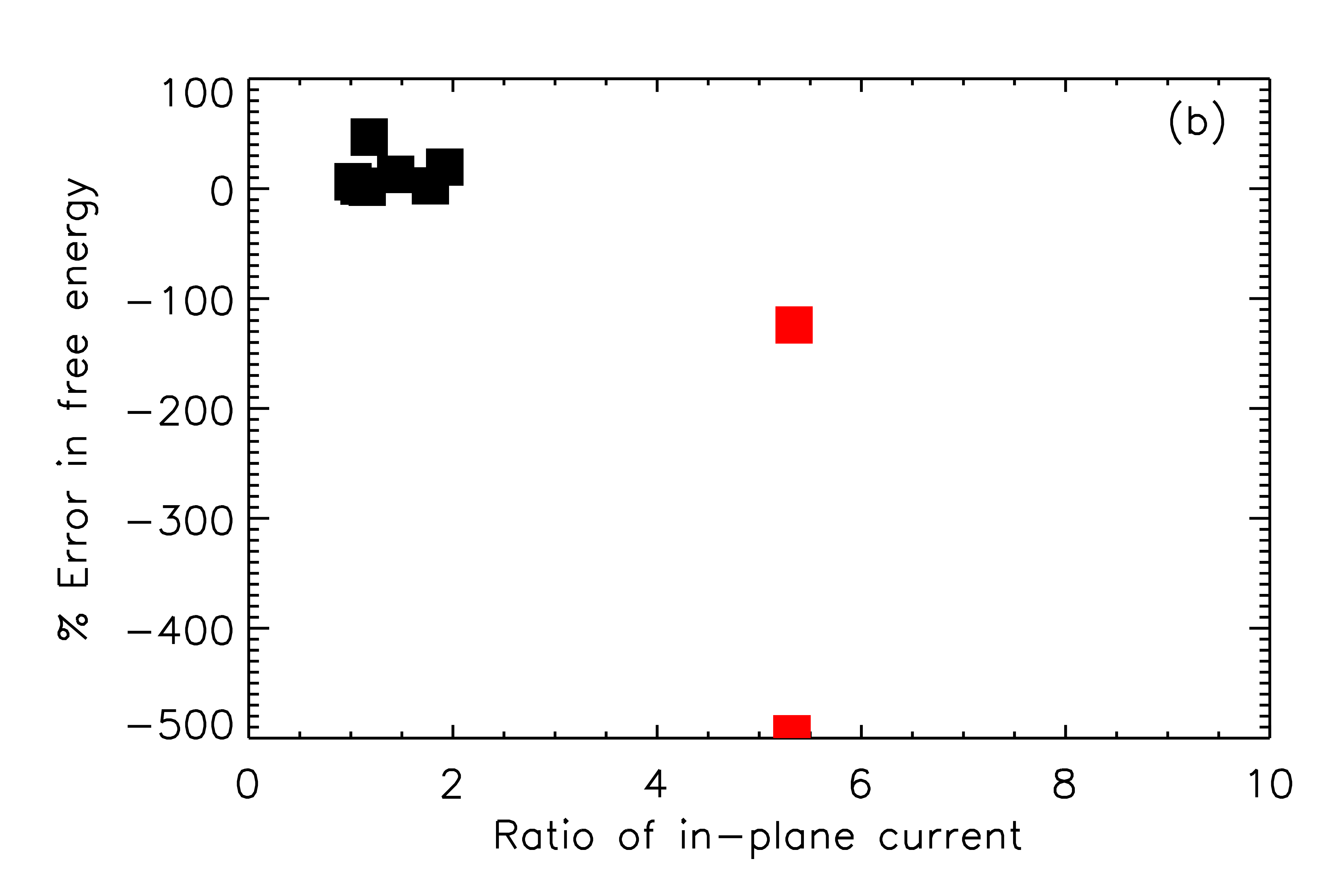}
\includegraphics[width=0.32\textwidth]{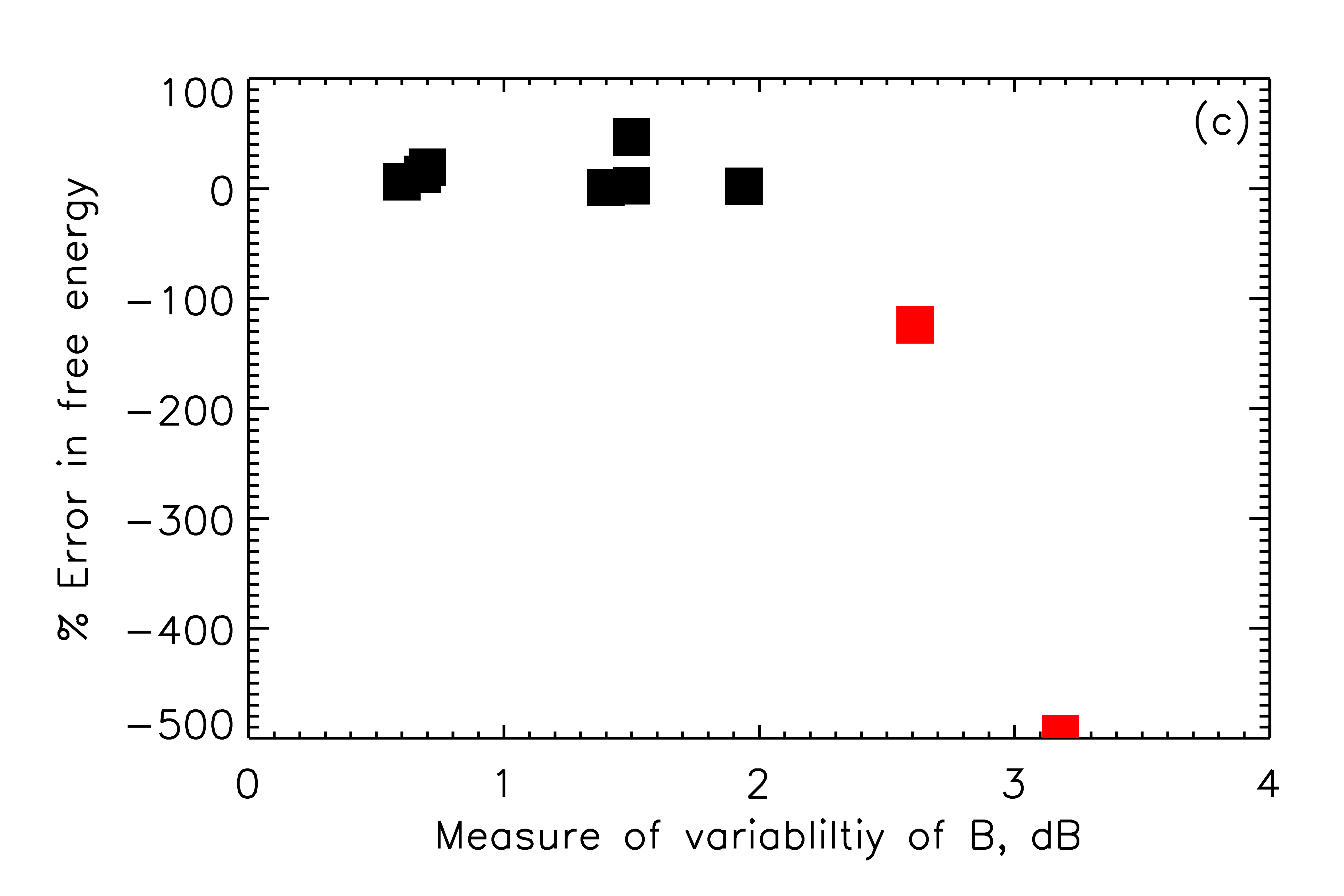}
\caption{ Panel (a): Error in free energy for the data-driven run vs the ratio of out-of-plane current (data-driven to ground-truth), for all the simulations in the parameter study. The red points are Simulations 5 and 7. Panel (b): The same but for in-plane current. Panel (c):  The same but for a measure for the variability of the vertical magnetic field data on a 12 minute cadence, $dB$.  \label{fig:error_current}}
\end{center}
\end{figure*}

Given that the free magnetic energy is related to the magnetic energy in the non-potential ($\mb{J}\ne0$) part of the magnetic field, it is worthwhile to analyze the amount of currents in the data-driven run compared to the ground-truth run. This is done for a simulation which shows small positive errors in free magnetic energy (Simulation 4), and one which has large negative errors (Simulation 5). The lower panels of Figures \ref{fig:currents1}-\ref{fig:currents2} show a measure of magnetic field variability $dB$ (red triangles) for the ground-truth data for Simulations 4 and 5, the error in free energy for the data-driven run (solid black line), and the ratio of out-of plane current between the data-driven and ground-truth run (solid red line). The ratio of out-of-plane current is given by $\sum{|J_{y}|_{driven}}/\sum{|J_{y}|_{ground}}$ with the sum over the domain above $z=0$. The variability measure $dB$ is a measure of how much the ground-truth magnetic field changes between driving input times of $dt=12$ minutes:
\be
dB(t) = \sum_{x_{i}}\left|{\frac{B_{z}(x_{i},t+dt)-B_{z}(x_{i},t)}{B_{z}(x_{i},t+dt)+B_{z}(x_{i},t)}}\right|
\ee
where $x_{i}$ are all points on the model surface ($z=0$). The upper panels in Figures \ref{fig:currents1}-\ref{fig:currents2} show the potential energy in the ground-truth simulation to highlight where in time during the emergence the errors occur.

As can be seen in Figure \ref{fig:currents1}, for Simulation 4, the variability measure $dB$ remains less than unity, the ratio of out-of-plane current remains close to unity, and the errors in free energy remain small ($<20 \%$). By contrast, in Simulation 5 (Figure \ref{fig:currents2}) the variability measure has periods where it is larger than unity, as does the ratio of current, and the free energy error gets more and more negative during the emergence.

Figure \ref{fig:error_current} shows plots of the free energy error against the ratio of out-of-plane current, the ratio of in-plane current $\sum{|(J_{x},0,J_{z})|}_{driven}/\sum{|(J_{x},0,J_{z})|}_{ground}$,  and magnetic field variability $dB$, for all nine Simulations. Those emergence events which exhibit large magnetic field variability (dB) over 12 minute intervals (5 and 7) have large currents generated in the data-driven runs and large, negative errors in the free magnetic energy. This is suggestive that the source of the large errors in the data-driven runs is the large variability of magnetic field and associated currents.

\begin{figure}[h]
\begin{center}
\includegraphics[width=0.49\textwidth]{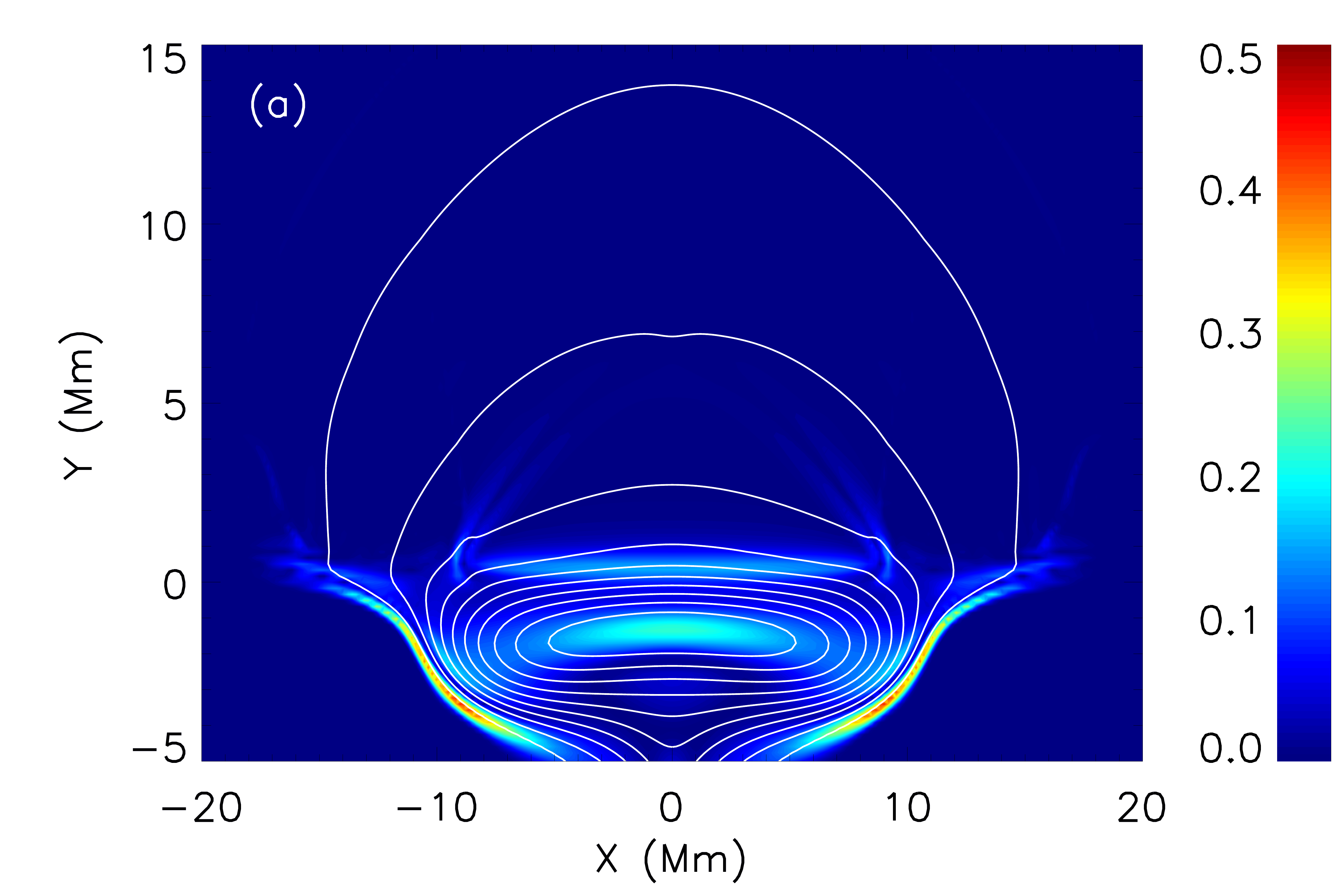}
\includegraphics[width=0.49\textwidth]{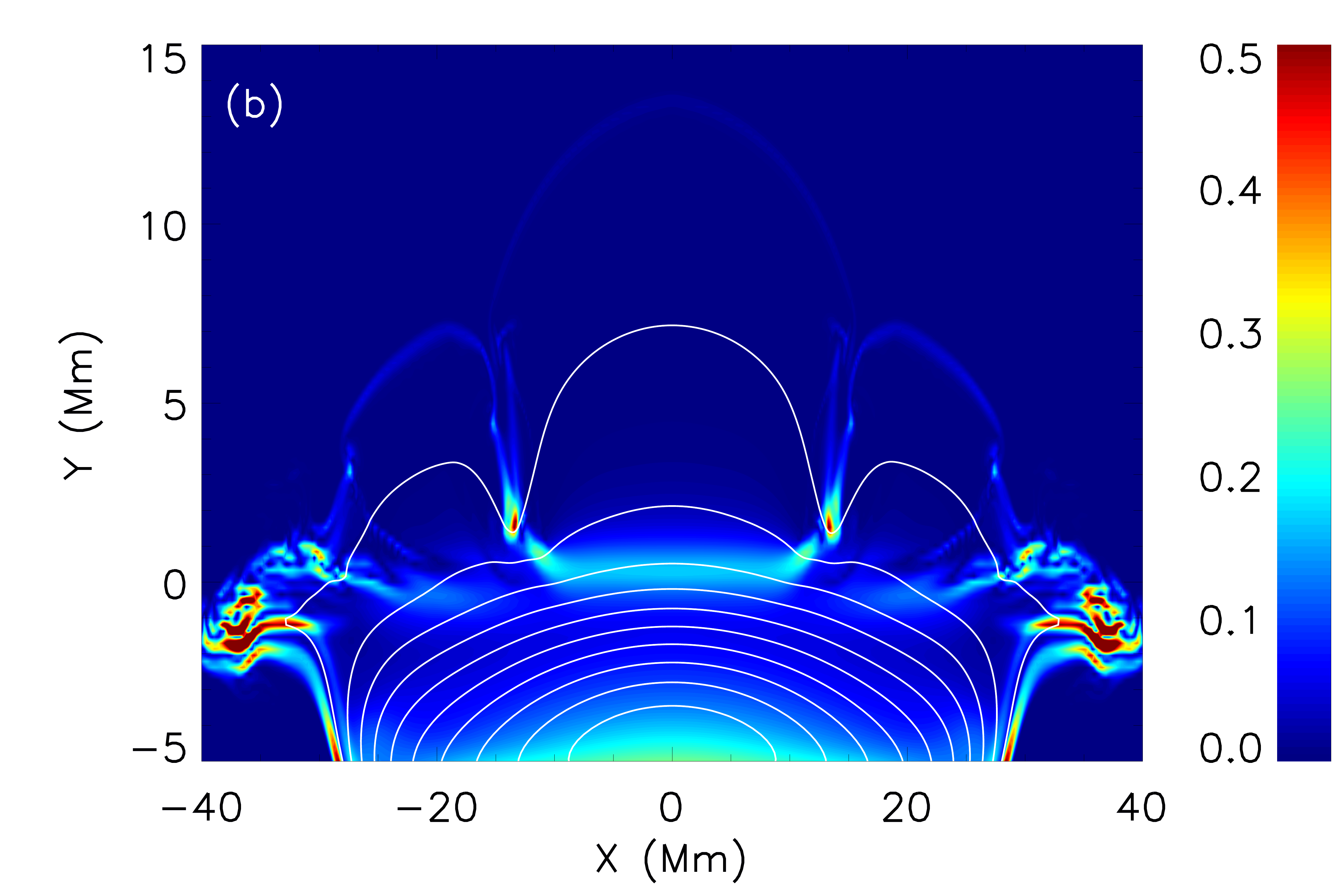}
\vspace{-10mm}
\caption{ Panel (a): Magnitude of current density (normalized by $B_{0}/\mu_{0}L_{0}$ where $B_{0}$ = 1300G and $L_{0}$=170km) and contours of flux $A_{y}$ (white lines) for Simulation 4 at 108 minutes. Panel (b): The same but for Simulation 5 at  264 minutes. \label{fig:Jz_Az}}
\end{center}
\end{figure}

To investigate the origin of these large currents in Simulations 5 and 7, the nature of the emergence, in particular the morphology of the emerging magnetic field, is examined. Figure \ref{fig:Jz_Az} shows the shape of the fieldlines in the 2D ($x,z$) plane in Simulations 4 (small positive errors in free magnetic energy) and 5 (large, negative errors). The fieldlines are represented by contours of constant flux $A_{y}$. Also shown are the magnitude of the current density in normalized units. Simulation 4 (panel a) shows a simple bipolar emergence, with no undulations of the fieldlines crossing the surface. Simulation 5 shows multiple undulations of fieldlines, and associated regions of larger current density. 

\begin{figure*}
\begin{center}
\includegraphics[width=0.49\textwidth]{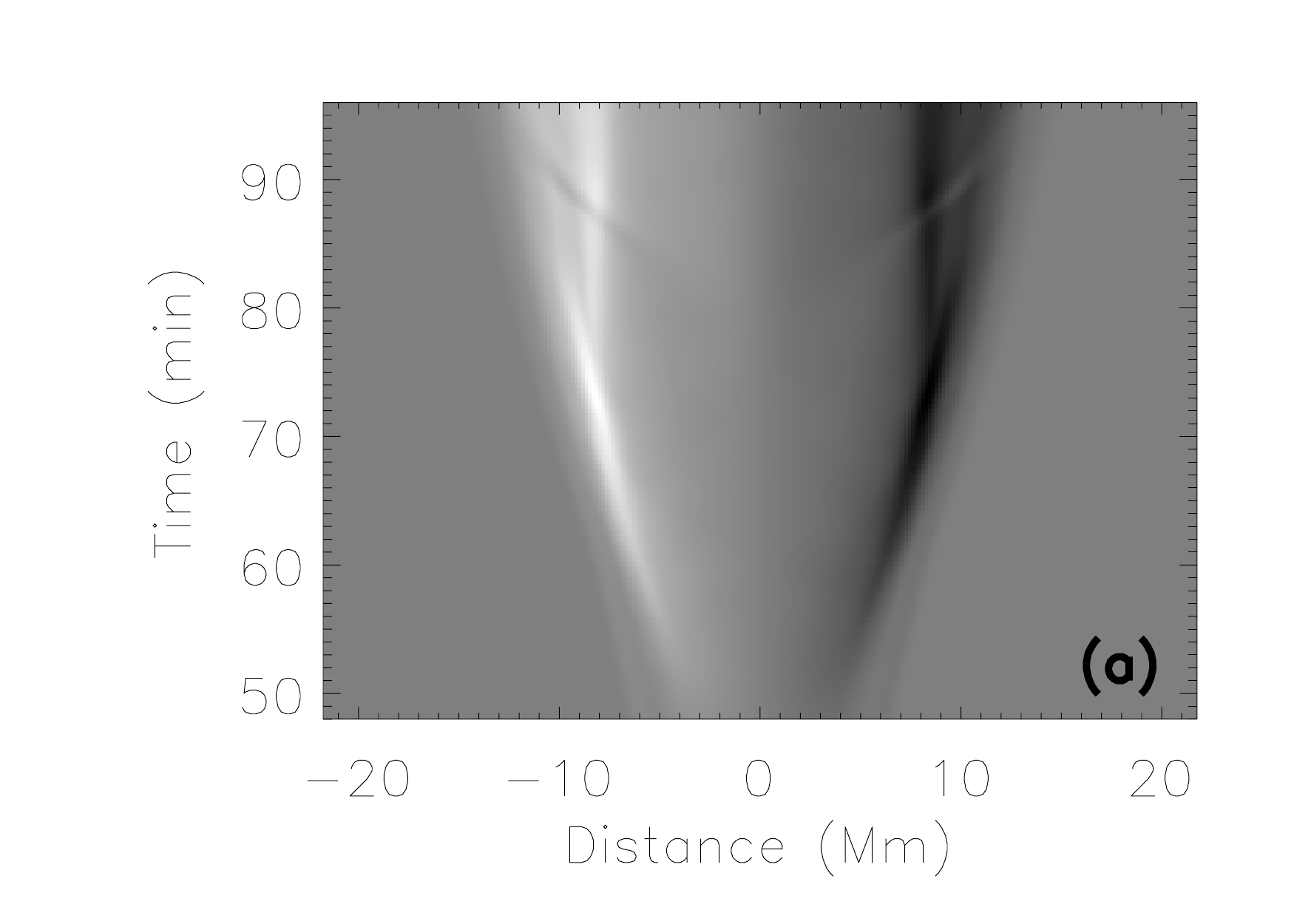} 
\includegraphics[width=0.49\textwidth]{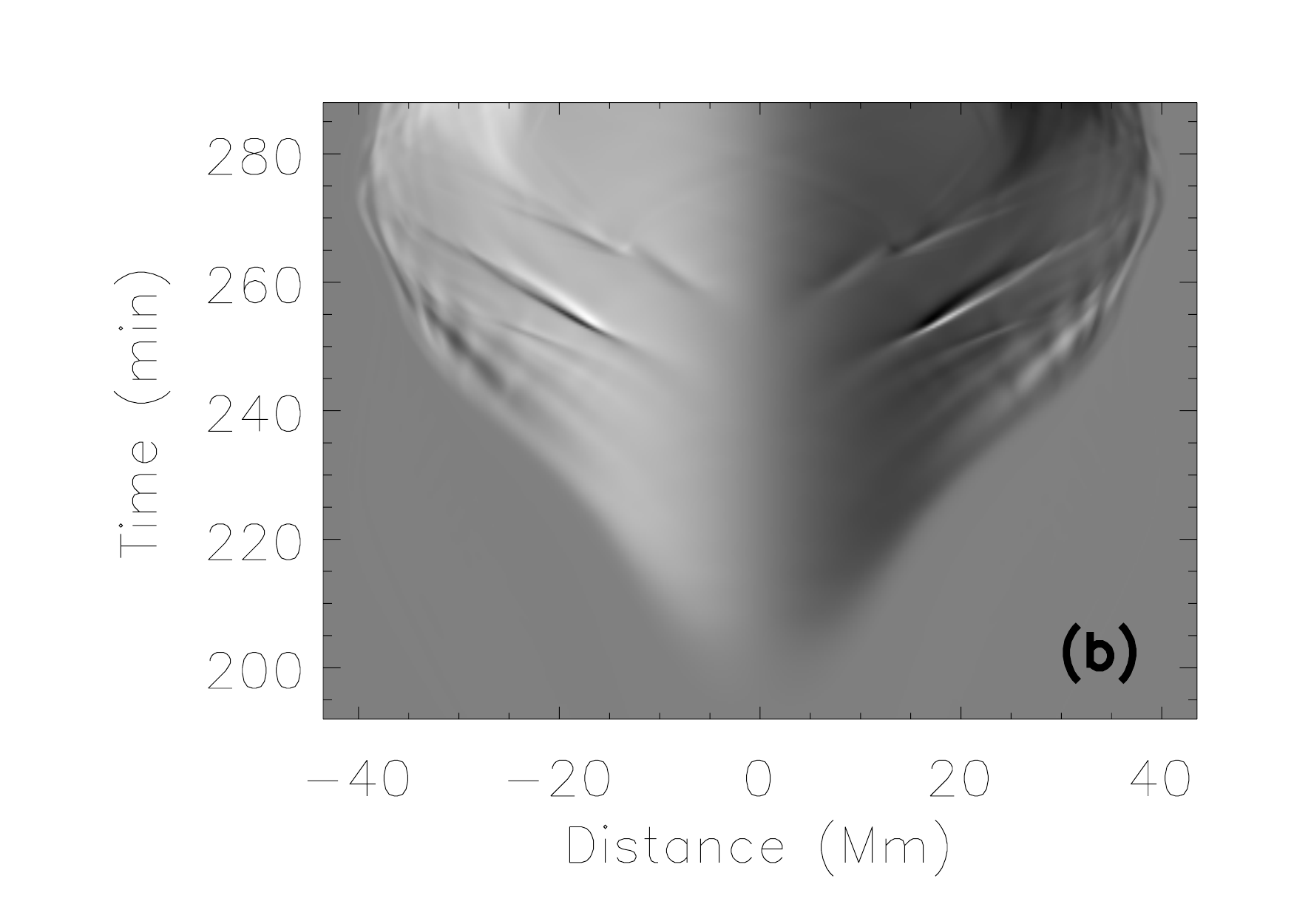}
\includegraphics[width=0.49\textwidth]{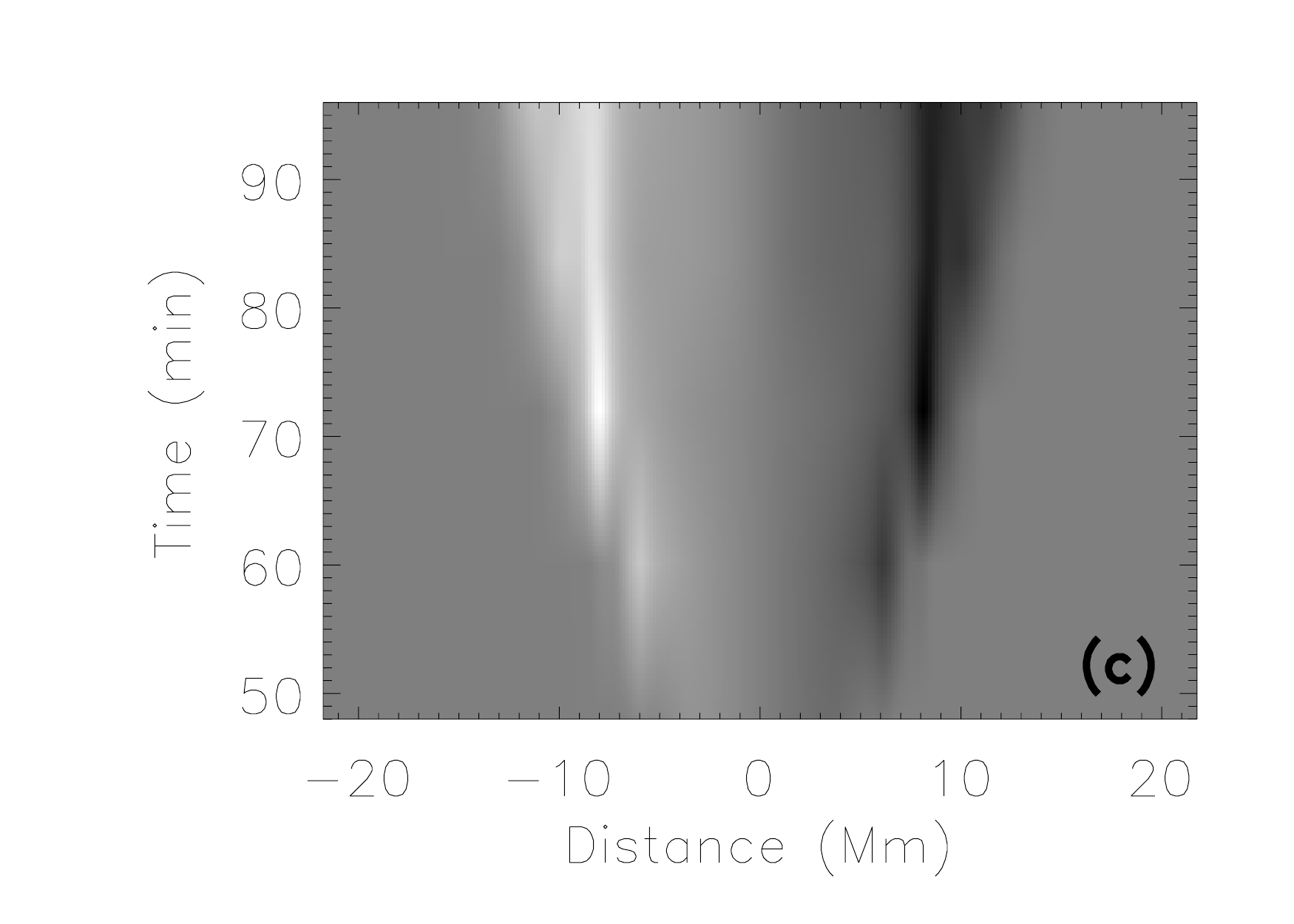} 
\includegraphics[width=0.49\textwidth]{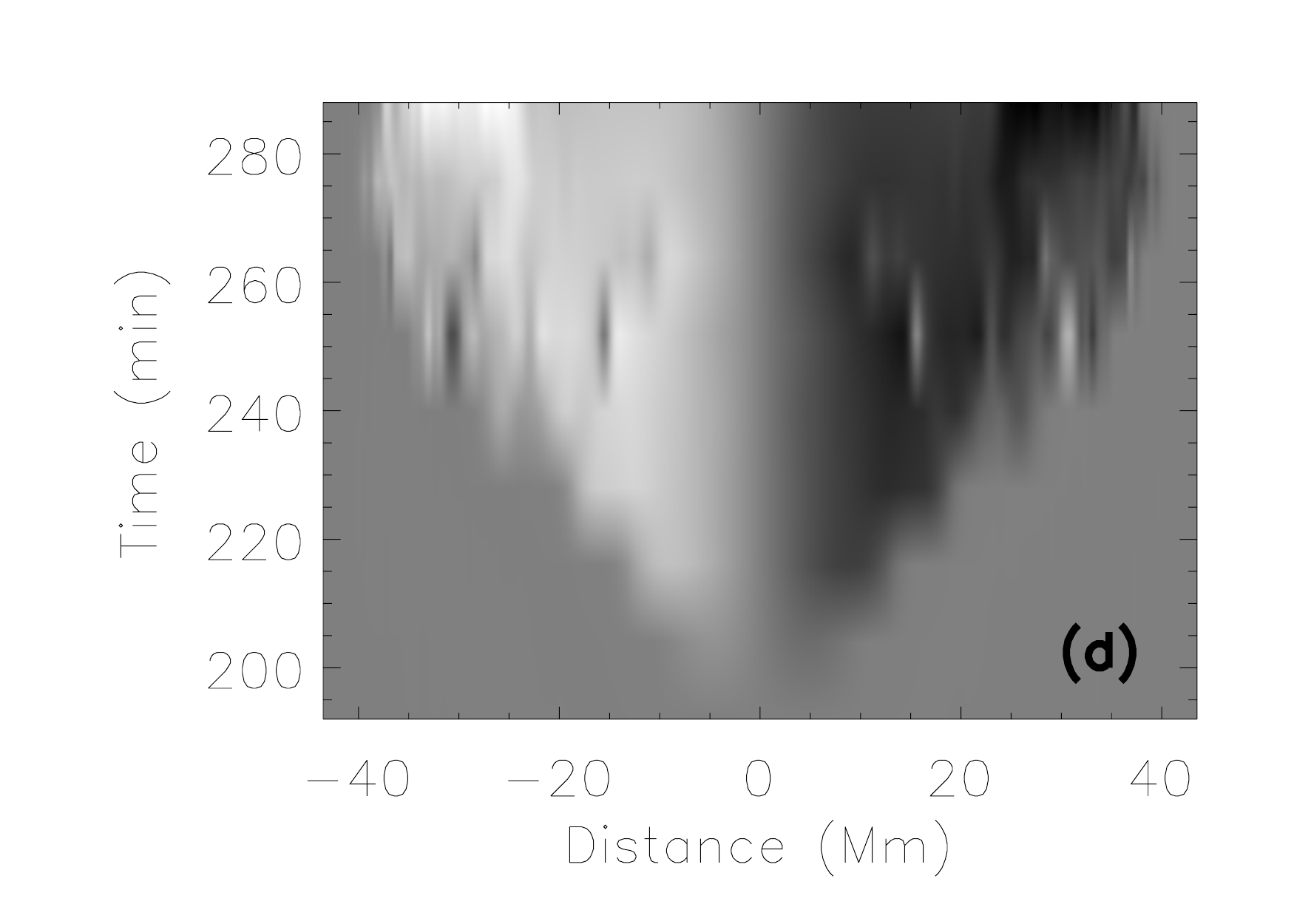} 
\caption{Panels (a) and (b):  Stacked plot of vertical magnetic field at the surface for Simulations 4 (a) and 5 (b). Panels (c) and (d):  The same but interpolated using 12 minute sampled data \label{fig:stackedB1}}
\end{center}
\end{figure*}

To represent these two emergence events in time, the top two panels of Figure \ref{fig:stackedB1} show the vertical field along the surface, stacked in time, for the ground-truth runs of Simulations 4 and 5 at a cadence of 24 seconds. For the simple emergence, e.g., as  that seen in Figure \ref{fig:Jz_Az}(a) for Simulation 4, one would expect two polarities slowly separating in time, as the magnetic flux associated with the original flux tube emerges though the surface. This is indeed seen for Simulation 4, Figure \ref{fig:stackedB1}(a).  However, for Simulation 5, shown in Figure 17(b), the signal of the vertical field is dominated by multiple pairs of  polarities which appear on each side of the neutral line of the active region and move outwards. A similar morphology occurs, but is not shown here, for Simulation 7 (the second simulation with large negative errors in the free magnetic energy in the data-driven run). The emergence associated with Simulations 5 and 7 is more complicated than a simple bipole; the magnetic field is undular and fieldlines cross the surface multiple times, resulting in multipolar regions forming on small timescales. 

The lower panels of Figure \ref{fig:stackedB1} show the same data as the top panels, but now sampled every 12 minutes, and interpolated to times (at a 24s cadence) between those 12 minute inputs. This is the data the data-driven run uses as a lower boundary condition. For Simulation 5 (panel d), with this interpolated data, the multipolar regions  result in regions where the magnetic field changes sign significantly over a 12 minute period. This explains why large surface magnetic field variability is seen in Simulations 5 and 7 ($dB>1$). In general, one would expect such variability and small scale structure, such as that seen in  Figure \ref{fig:stackedB1}(b), in real active regions. 

It is hypothesized here that this variability is the source of the spurious currents in the data-driven runs for Simulations 5 {\jl and} 7, when the driving interval violates a sampling condition associated with the apparent motion of these small-scale bipoles along the photosphere. If $\tau$ is defined as the timestep at which the photospheric data is sampled, the apparent velocity of the small-scale bipoles is defined as $v_{h}$, and their spatial extent is characterized by $L$, then the condition for the sampling interval  is $\tau < dt_{sample}\equiv L/v_{h}$. If the condition is violated, then between sampling times, the bipoles move a distance $\tau v_{h}> L $, i.e., a distance larger than their spatial extent, and this under-sampling thereby effectively generates a strobe effect in the driving data, where the bipoles appear to jump across the photosphere, rather than move continuously. 
Note, that this condition is different from the CFL limitation of the numerical integration, $dt_{CFL} \equiv dx / \textrm{max}(|\mb{V}|, C_{s}, C_{A})$

 For the ground-truth dataset for Simulation 5 (seen at a cadence of 24s in panel (b) of Figure \ref{fig:stackedB1}), the spatial extent of the bipoles (half-width half-max of one polarity of the bipole) is approximately $L=$1 Mm and the apparent horizontal velocity is $v_{h}$=20 km/s, which requires a sampling interval less than $dt_{sample} = L/v_{h}=$ 50 s. The violation of this constraint, when the data-driven run uses photospheric data sampled every 12 minutes, is manifest in  Figure \ref{fig:stackedB1}(d), where one can see numerous small scale features appearing and disappearing in the pseudo-magnetograms, when sampled every 12 minutes,
 due to the strobe effect. To reconcile this time dependent boundary condition, which has rapidly appearing and disappearing bipoles, the data-driven run generates large electric currents, which propagates into the domain. This results, as was shown in Figure \ref{fig:currents2}, in spurious electric currents in the data-driven model atmosphere, and a larger free magnetic energy, compared to the ground-truth run.


\begin{figure}[h]
\begin{center}
\includegraphics[width=0.4\textwidth]{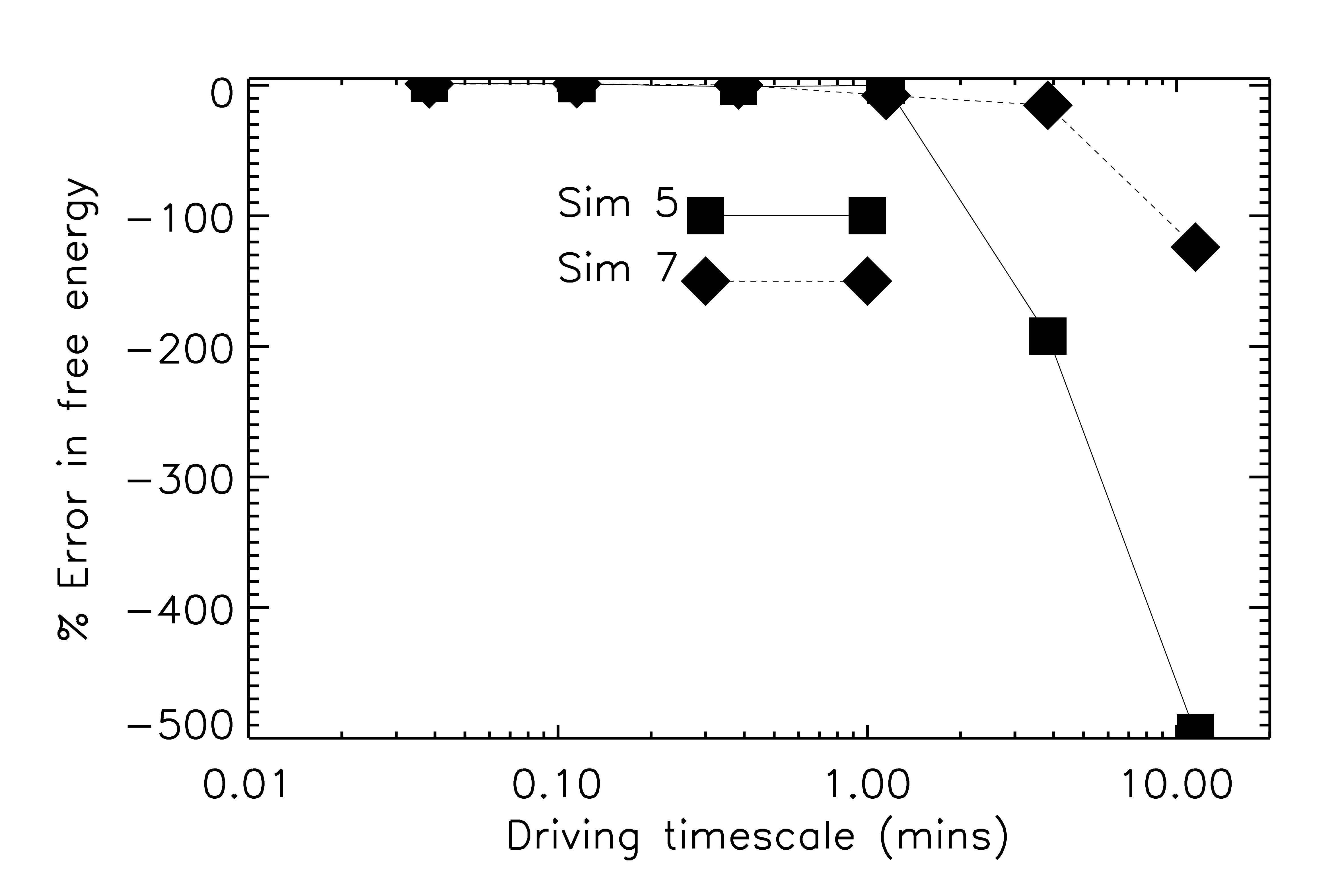}
\caption{\% Error in free magnetic energy for Simulations 5 and 7, as a function of driving timescale. For both these simulations, a driving cadence of $\sim 1$ minute is required to obtain errors with magnitude less than 10\%  in the data-driven run.\label{fig:bad_sims_cadence}}
\end{center}
\end{figure}

To highlight the importance of the driving timescale on the accuracy of the data-driven model for these multipolar emergence events, Figure \ref{fig:bad_sims_cadence} shows the  error in the free magnetic energy in the data-driven runs for Simulations 5 and 7 as a function of the driving timescale. At the maximum driving timescale of 12 minutes, the errors are large and negative, as the large variability of the magnetic field in these multipolar emergence events cannot be captured on such a timescale (as shown above). However, as the driving timescale is reduced to about a minute, the errors in the data-driven run reduce to less than a few percent, as the driving timescale is now able to capture the magnetic field variability related to the undular, multipolar emergence. This is consistent with the estimate of the required driving interval of 50 s from above.

\section{Discussion}

This study aimed to assess the accuracy and feasibility of calculating the coronal magnetic field above active regions using dynamic data-driven MHD simulations with data provided at the photosphere. As there currently exist no observations that can provide us with a ground-truth dataset for the magnetic field in the corona above the active region to test the data-driven MHD runs, we created multiple test cases using nine different dynamic MHD simulations of flux emergence from below the surface. Then we used the data at the surface from these ground-truth runs as synthetic magnetograms to drive data-driven runs of the atmosphere above the surface. To simplify our study, we assumed that all MHD variables are known at the surface, not just the magnetic field. This is not the case for current observations of the photosphere, where only vector magnetic field observations are reliably observed at high cadence over full active region scales.  
We then investigated the effect of the driving timescale on the errors produced in the coronal magnetic field solution for the data driven runs. We investigated timescales across the range from the CFL limited timescale ($dt_{CFL} \sim $ 0.004 min) all the way up to the 12 minute timescale typical of HMI observations. Furthermore, we conducted a parameter study by varying the model parameters in our ground-truth simulations to investigate the effect of the emergence timescale on the errors. 

We  found that, at a given emergence timescale, the error in the coronal magnetic field decreased with decreasing driving interval, as expected, and that at the fundamental (sub-CFL) timescale of the numerical method, we were able to recover the ground-truth solution. Furthermore, we found that at the standard HMI interval of 12 minutes, the error in the coronal magnetic field solution decreased with emergence timescale, with the error in the free magnetic energy being as low as 1\% for the active region that emerged over a day. This is encouraging for the future development of data-driven MHD models which use photospheric magnetic field measurements, and derived values for the other MHD variables,  to model the dynamic evolution of the coronal magnetic field. 

However, we also found exceptions to the above results, where large negative errors in the free magnetic energy were found in the data-driven solutions. Preliminary analysis indicates that this is due to the emergence and evolution of multiple sub-AR scale bipolar regions which evolve on small timescales, and are thus under-sampled by the 12 minute interval driving data. The data-driven runs of these undular, multipolar regions generated large electric currents which created spuriously large free energy in the atmosphere. We hypothesized that these currents were a result of the strobe effect generated by features moving at a rate faster than can be captured by the input cadence.

We defined a critical sampling condition $dt_{sample}=L/v_{h}$ for the data-driven runs in these complex emergence events, where $L$ and $v_h$ are the characteristic size and apparent velocity of
the sub-AR scale bipolar regions. For the parameters we have considered, this sampling condition was approximately one minute. By reducing the driving timestep down to, and below, this critical value, we resolved the problem of spurious currents and free magnetic energy in the data-driven runs, and we recovered errors in the magnetic free energy (compared to the ground-truth) below a few percent. However, when using real photospheric data, an effort is clearly necessary to calculate this condition for the active region being studied.
 This study shows that extreme care must be taken when using solutions of magnetic field from data-driven MHD models of complex emerging active regions which contain rapidly evolving magnetic features. Fortunately, the potential availability of higher cadence magnetic field measurements (90-135 s) from HMI-data \citep{Xudong} will allow for more analysis.

It is important to note that  this investigation assumes the best-case scenario, a scenario that is required by MHD, that \textit{all} the MHD variables (including velocity, pressure and temperature), not just the magnetic field, are provided without error at the given cadence. This is currently beyond the state of the art in the observations, {\jl where, for example, the velocity must be inferred from the time-series of the 
photospheric magnetic field. The errors investigated in this study, which come from the lack of temporal information at the photosphere, will always be present in data-driven models regardless of the methods used to mitigate the lack of information about the velocity and temperature. Future investigations will study the introduction of errors in data-driven coronal field models due to (1) this lack of information at the photosphere and methods used to mitigate it, (2) reduced spatial cadence of photospheric data, and (3) instrument noise and bias.}


\begin{acknowledgments}
\nind{Acknowledgements:}
This work has been supported by the NASA Living With a Star \& Solar and Heliospheric 
Physics programs, and the Chief of Naval Research 6.1 Program. 
The simulations were performed under a grant of computer time from the DoD High Performance Computing program. {\jl We thank the anonymous referee for useful comments on the manuscript.}
\end{acknowledgments}

\bibliography{bib2}

\newpage

\end{document}